\documentclass[10pt,onecolumn]{article} 

\usepackage[a4paper,includeheadfoot,margin=1cm]{geometry}
\usepackage{mathptmx}
\usepackage{txfonts}

\usepackage[T1]{fontenc}
\usepackage{ae,aecompl}
\usepackage{times}
\usepackage{amsmath}
\usepackage{graphicx}
\usepackage{amssymb}
\usepackage{url,hyperref}
\usepackage{listings}
\usepackage{natbib}
\usepackage{array}

\newcommand\aap{A\&A}                
\newcommand\aj{AJ}                   
\newcommand\apj{ApJ}                 
\newcommand\apjl{ApJ}                
\newcommand\apjs{ApJS}               
\newcommand\araa{ARA\&A}             
\newcommand\mnras{MNRAS}             
\newcommand\nat{Nature}              

\hypersetup{colorlinks=true,linkcolor=blue,citecolor=blue,filecolor=blue,urlcolor=blue,}
\begin{document}
\title{Photo-evaporation of proto-planetary gas discs due to flybys of external single stars in different orbits}

\author{Dai Yuan-Zhe$^{1}$,
	Liu Hui-Gen$^{1*}$,
	Wu Wen-bo$^{2}$,
	Xie Ji-Wei$^{1}$,
	Yang Ming$^{1}$, 
	Zhang Hui$^{1}$,
	Zhou Ji-Lin$^{1}$\\
$^{1}$School of Astronomy and Space Science, Nanjing University, 163 Xianlin Avenue, Nanjing 210046, China \\
$^{2}$ Key laboratory of optical astronomy, National Astronomical Observatories, Chinese Academy of Sciences,\\
20A Datun Road, Chaoyang District, Beijing 100012, China
\\
\\
huigen@nju.edu.cn \\
}

\maketitle
\thispagestyle{empty}

\begin{abstract}
During the evolution of proto-planetary disc, photo-evaporations of both central and external stars play important roles. Considering the complicated radiation surroundings in the clusters, where the star formed, the proto-planetary discs survive in different lifetimes due to flyby events. In this paper, we mainly focus on the disc around a T Tauri star, which encounters with another main-sequence star with different temperatures in hyperbolic orbits with different peri-center distances, eccentricities and inclinations. We find the criterion for gap-opening due to photo-evaporation of central star after the flyby event. A gap is opened in the late stage of gas disc, and induce that the gap only influence the planet formation and migration limitedly. If the flyby orbit has a moderate value of peri-center distance, which weakly depends on the eccentricity and inclination, the external photo-evaporation lead to a maximum mass loss during the flyby event. Flyby stars in orbits with smaller eccentricities or larger inclinations induce larger mass loss. Adopting a simple multiple flyby models, we conclude that in open clusters, gas discs usually survive in typical lifetimes between 1 and 10 Myr, except there are many massive stars in dense open clusters. In globular clusters, discs disperse very quickly and hardly produce the gas giant planets. The fast-depleted discs are probably responsible for the null detection of giant planets in globular clusters.

\end{abstract}

\section{Introduction}
The environments of star formation region are usually clustered. Currently, only two pulsars as planet hosts are detected in globular clusters and tens of planet hosts in open clusters. Most recent works by
\citet{2017Natur.548..183M}, indicate the number of free-floating or wide-orbit Jupiter-mass planets in the galactic halo, i.e. less than 0.25 Jupiter per main sequence star on average. Only eight giant planets are detected around stars in clusters, while seven of them are in open clusters \citep{2018aJ....155..173C,1999apJ...523..763T}. Because of the limited planet sample in open clusters, there are debates whether the planet occurrence rate around stars in open clusters and field stars are similar or not. However, some transiting surveys in globular clusters with null results hint the occurrence rate of Jupiter-mass planets in clusters may be lower than field stars \citep{2000apJ...545L..47G,2005apJ...620.1043W,2012a&a...541a.144N,2017aJ....153..187M}

The instability of planets in clusters is considered to be an important role \citep{2009apJ...697..458S,2013apJ...772..142L}. A large fraction of planets very close to the host star are probably stable in open clusters or even in the outer region of globular clusters \citep{2011MNRaS.411..859M,2001MNRaS.322..859B}. Beside the dynamical instability in clusters, the metallicity is also considered to influence planet formation \citep{2012Natur.486..375B}. Low metallicity in old globular clusters may lead to lower planet formation rate, because the few dust around stars with poor metallicity can hardly form enough planetary embryos. Additionally, the metallicity is crucial for the core due to gas accretion scenarios, but once the core is formed before gas depletion, the gas giant can form. 


Due to core accretion scenarios, the formation of giant planet is relative to the proto-planetary discs tightly. The mass loss and evolution of proto-planetary gaseous disc play important roles during planet formation, especially for gas giants. In viscous discs, accretion from central star make the density profile decays smoothly, and lead to disc dispersion. Photo-evaporation is also an important mechanism to influence the evolution of gaseous disc. The evolution, especially the lifetime, of the gaseous disc will determine weather the gas giant can form or not, and how large the gas giants can grow. 

As one of essential mechanisms to lose mass, photo-evaporation leads to dispersal of proto-planetary discs \citep{2011aRa&a..49...67W}. Photo-evaporation can be driven by photons in the energy range of far-UV( 6 eV<$h\nu$<13.6 eV, here after FUV), extreme-UV(13.6 eV<$h\nu$<0.1 keV,  here after EUV), and X-ray($h\nu$>0.1 keV). Photons in such energy ranges influence the discs in different ways. FUV can dissociate molecules like $H_{2}$, while EUV and X-ray can ionize the hydrogen atoms. The proto-planetary disc is heated by these photons to temperatures about 1000-10000 K, and consequently gas materials is evaporated via thermal escape. 

Previous photo-evaporation models of are usually considering EUV, FUV and X-ray in a viscous disc. Most of early models focused on the radiation from central stars \citep{2011aRa&a..49...67W}. \citet{2001MNRAS.328..485C} first introduce a UV-switch model which includes the internal EUV photo-evaporation and disc viscosity. They indicate a photo-evaporative gap in the late time of disc evolution. With a mass loss rate about $10^{-10}$ M$_{\odot}$ yr$^{-1}$ originated by central ionizing flux, at the early stage in the disc evolution, the internal photo-evaporation has a negligible effect compared with the large accretion rate due the viscosity. As the disc disperses, the internal photo-evaporation becomes crucial and clear out the gas material near the gravitational radius while the viscous evolution can hardly fill the gap. Then \citet{2006MNRAS.369..216A,2006MNRAS.369..229A} develop the model and consider the effect of stellar radiation directly incident on the inner disc edge at late disc times which lead to a quick dispersion of the out disc when the gap is opened.  
Recent photo-evaporation models include internal X-ray \citep{2009ApJ...699.1639E,2010MNRAS.401.1415O,2011MNRAS.412...13O} or FUV irradiation \citep{2009ApJ...690.1539G}, in addition to EUV photons. It's because that X-ray and FUV photons can penetrate the larger columns of neutral gas than EUV photons and consequently lead to larger mass loss rate $\sim 10^{-8}$ M$_{\odot}$ yr$^{-1}$ which is two order magnitude larger than that driven by EUV photons. To speak more specifically, the FUV photons can remove disc mass at large radii, and truncate the disc at $\sim 100$ AU typically. X-ray photons can penetrate the neutral gas to a few tens AU. While EUV photons can only penetrate neutral gas to several AU efficiently. I.e. the mass loss rate due to EUV photons is more concentrated at a scale of a few AU.  

The external photo-evaporation is also important during disc evolutions. \citet{1998apJ...499..758J} presented a model for the photo-evaporation of circumstellar discs or dense clumps of gas by a massive external source of ultraviolet radiation. Later, in 2003, they developed a model in the proto-planetary discs around T Tauri stars \citep{2003ApJ...582..893M}. They combined viscous evolution, central EUV photo-evaporation and external EUV or FUV photo-evaporation originated by massive nearby stars. External photo-evaporation will shrink the disc outer edge and accelerate the disc evolution. Photo-evaporation will not only influence the lifetime of proto-planetary disc, but also influence the internal structure of the disc. Similar to internal photo-evaporation models, at late times, there will be a photo-evaporative gap. The gap structure will influence both the runaway accretion of gas giant and the migration of small planets. Thus planet formation and orbital architectures may be various during the disc evolution under photo-evaporation. 


In this paper, we mainly focus on the gaseous disc around a T Tauri star,  which undergoes a close encounter with another main-sequence star. The photo-evaporation effect depends on the orbit parameters of the intruder in a hyperbolic orbit. Previous external photo-evaporation models usually assume that the radiation field is perpendicular to the disc surface, while it's not suitable in the condition of flyby events with random orientation. We combine both orbit parameters and efficient receiving area of disc during the flyby, then develop the photo-evaporation model of external stars to a general way. Besides, the duration of a close encounter is usually accounted until the star is too far to influence the mass lose of gaseous disc. It's our aims to figure out the influence on the viscous disc due to flybys with different stars in different orbits. Furthermore, multiple flyby events will influence the disc evolution continuously, especially in the environments of clusters, where the stars formed. We focus on the lifetime of gaseous disc in different environments (e.g. field stars, open clusters and globular clusters), and provide some clues of gas giant formation in these environments according to different disc lifetime.



In section 2, We developed an alternate EUV/FUV photo-evaporation model of external star due to different orbital parameters.  In section 3, we analysis the relation between the hyperbolic orbit parameters and mass loss, via integration. In section 4, the one-dimension alpha disc simulations are represented and be compared with the analysis results of photo-evaporation. In section 5, our results are extended in young open clusters and globular clusters. Additionally, we compare our results with previous works and observations in clusters. We summarize our conclusions in section 6. 

\section{The photo-evaporation model}
The evolution and dispersal of the irradiated proto-planetary discs are controlled by the viscosity of disc and photo-evaporation originated from both central star and external star. The well-known diffusion equation \citep{1974MNRaS.168..603L,1981aRa&a..19..137P} describes the viscous evolution of an one dimension disc. Additionally, \citet{2001MNRAS.328..485C} considers both viscosity and photo-evaporation simultaneously, and describes the evolution of gaseous disc as follows:
\begin{equation} \label{1}
\frac{\partial \Sigma}{\partial t} = \frac{3}{r}\frac{\partial}{\partial r}\left[r^{1/2}
\frac{\partial}{\partial r}\left( \nu(r) r^{1/2} \Sigma \right)\right ]-\dot{\Sigma}_{\rm  w}\left(r,t\right), 
\end{equation}
where $\Sigma$ is the surface density of disc, $\nu(r)\propto r$ is the viscosity of the disc, and $\dot{\Sigma}_{\rm  w}\left(r,t\right)$ is the surface density loss caused by central star. Obviously, the photo-evaporation of the external star influence the disc mass loss during disc evolution. In our model we rewrite the equation in order to consider the photo-evaporation from both the central star and the external star, which is
\begin{equation} \label{2}
\frac{\partial \Sigma}{\partial t} = \frac{3}{r}\frac{\partial}{\partial r}\left[r^{1/2}
\frac{\partial}{\partial r}\left( \nu(r) r^{1/2} \Sigma \right)\right]-\dot{\Sigma}_{\rm  c}-\dot{\Sigma}_{\rm  ex},
\end{equation}
where $-\dot{\Sigma}_{\rm  c}$ is the surface density decayed rate caused by central EUV photo-evaporation, while $-\dot{\Sigma}_{\rm  ex}$ is the decayed rate originated from external FUV or EUV photo-evaporation. The formula of the $\dot{\Sigma}_{\rm  c}$ and $\dot{\Sigma}_{\rm  ex}$ is given in \citet{2003ApJ...582..893M}( Equation (14)(17) and (19) in their paper). 

The luminosity from the center is composed of two parts: the luminosity ($L_{\rm  c}$) from the central star and the luminosity $L_{\rm  d}$ from the inner accretion disc. Considering that $L_{\rm  d}/L_{\rm  c}\gtrsim 10^{-3}$\citep{2011aRa&a..49...67W}, in the lifetime of the inner disc, the luminosity from the central star dominate the total luminosity. Considering the accretion of disc material, the ionization luminosity of the host star is nearly the constant in the time scale of 1 million years \citep{2003ApJ...582..893M}.  Therefore, We don't consider the change of ionization luminosity of host star duration the longevity of the disc. Besides, We only take the central star's EUV photo-evaporation into consideration. There is a gravitation radius $r_{\rm  g,euv}$, where the sum of kinetic energy and thermal energy of heated gas equals to the gravitational potential. EUV photo-evaporation dominates the outer area where $r>r_{\rm  g,euv}$, and the gravity bounded the heated gas in the inner area where $r<r_{\rm  g,euv}$. Thus, the term $\dot{\Sigma}_{\rm  c}$ adopted in this paper is:
\begin{equation} \label{3}
\dot{\Sigma_{\rm  c}}\left(r \right)=
\begin{cases}
2.75 \times 10^{-12} \rm M_{\rm  \odot}\cdot \rm {yr^{-1}} \cdot \rm {AU}^{-2} \left(\frac{\Phi_{\rm  c}}{10^{40} s^{-1}}\right)^{1/2} \\  \left(\frac{\beta r_{\rm  g,euv}}{\rm 1AU}\right)^{-1.5} \left(\frac{r}{\beta r_{\rm  g,euv}}\right)^{-5/2}, &r\geq \beta r_{\rm  g,euv} \\0, &\rm otherwise.
\end{cases}
\end{equation}
where $\Phi_{\rm  c}$ is the ionization luminosity of the central star, and $\beta < 1$ is the soften factor. Theoretically, inside the gravitationally radius $r_{\rm g,euv}$, gas material has lower energy to escape from the gravity. However, there is mass loss inside $r_{\rm g,euv}$. As the result of the net inner flow of the gas, the accurate gravitational radius will shrink due to this extra pressure gradient, thus we assume the effective gravitational radius $\beta$ $r_{\rm g,euv}$, where $\beta=0.5$ according to \citet{2003ApJ...582..893M}.  $r_{\rm  g,euv}$ is set as the gravitational radius where the gas temperature is $\sim 10^{4} $ K. $\Phi_{\rm  c}$ depends on the stellar parameters, yet we fix it as 10$^{41}$ photons s$^{-1}$ \citep{2001MNRAS.328..485C}. 

We mainly focused on the dynamical and thermal mechanism of the external photo-evaporation. The energy of the FUV photons is lower than ionizing energy of hydrogen atom, while it can penetrate and heat the neutral gas. In EUV-dominated flows, thermal pressure at the disc surface is dominated by photo-ionization, and the FUV produced photo-dissociation region (PDR) remains relative thin, FUV photons penetrate this thin region and heat the gas to subsonic, so the mass loss rate is controlled by EUV photons \citep{1998apJ...499..758J}. In FUV-dominated flows, the FUV photons heat the surface of the disc, and thermal neutral gas expand and reach the edge of the disc as warm wind.

In our model, we assume that most of the material is removed at the disc edge, $r_{\rm d}$, and photoevaporation from the disc can be approximately by photoevaporation from a sphere with radius $r_{\rm d}$\citep{2003ApJ...582..893M}. Yet they didn't consider induced angle of the radiation field which will affect the effectively radiation area. In their case, the direction of the nearby radiation field is perpendicular to the disc, which is not a suitable assumption in a more complicated surroundings such as in the core of young open clusters or GCs. Besides, they consider a constant distance $d$ between two stars. Actually, the photons flux is proportional to $d^{-2}$. In the case of flyby, $d$ must vary with time. So we developed the formula of the external photo-evaporation including induced angle of the radiation (which is the function of orbit parameters) and distance (which is also the function of orbit parameters).

\begin{figure*} 
	\centering
	\includegraphics[width=0.9\linewidth]{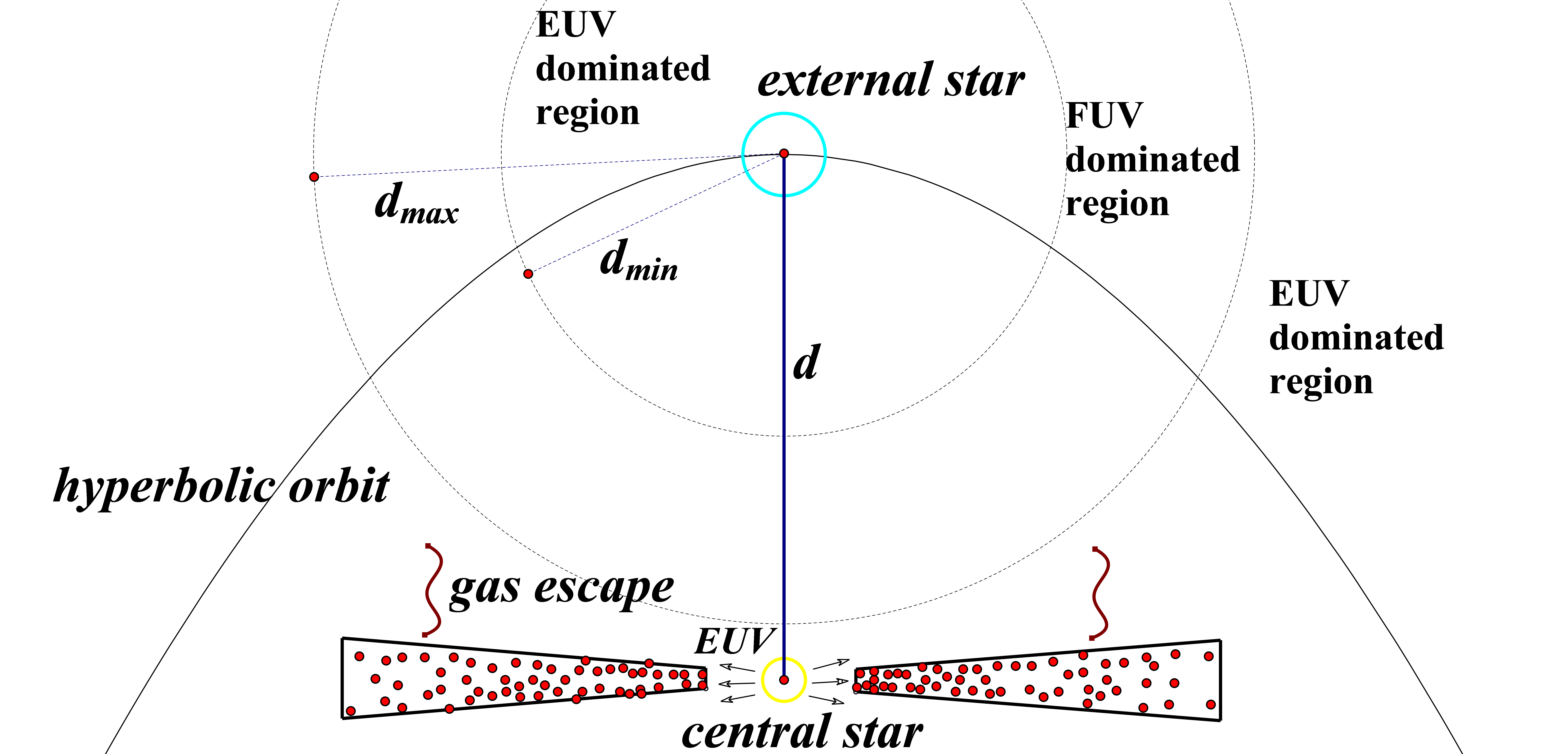}
	\caption{The sketch map of the photo-evaporation model. It shows a disc around a T Tauri Star when it is going through a close encounter with another star, with the combination effect of central and external photo-evaporation. The gas in the disc will escape due to thermal escape. If the distance $d$<$d_{\rm  min}$ or $d>d_{\rm  min}$, the EUV from external star dominates photo-evaporation, thus the photo-evaporation rate can be expressed as Equation \ref{7}. When $d_{\rm  min}<d<d_{\rm  max}$, the FUV from external star dominates photo-evaporation, and the photo-evaporation rate can be expressed as Equation \ref{10}. \label{Figure 1}}
\end{figure*}

The FUV or EUV photons of external stars will dominate the photo-evaporation, depending on the distance of the external stars \citep{1998apJ...499..758J}. Figure \ref{Figure 1} shows a sketch map of photo-evaporation during a flyby events.  When the external source is far away from the central star, the FUV photons can not heat the gas in the disc surface to about $\sim 1000$ K and result in relatively thin PDR, in which the column density is $\sim 10^{20}$ cm$^{-2}$. In other words, EUV photons will penetrate the thin PDR, and heat the disc surface, because the energy of EUV photons is higher than the FUV photons. While the external source is close to the central star, the EUV radiation field is extremely strong and its easy to penetrate the PDR, eventually evaporates the gas. Thus, there is only a narrow region dominated by FUV photons. To distinguish the EUV and FUV dominated region, there are two distances, $d_{\rm  min}$ and $d_{\rm  max}$, proposed by \cite{1998apJ...499..758J}. The minimum distance can be decided as follows:
\begin{equation} \label{4}
d_{\rm  min} \approx 5 \times 10^{17} \left( \frac{\epsilon^{2}}{f_{\rm  r}\Phi_{\rm  49}} \right)^{-1/2} r_{\rm  d14}^{1/2} \rm cm.
\end{equation}
where 
\begin{equation} \label{5}
\epsilon=\left(\frac{N_{\rm  D}}{10^{21}\rm cm^{-2}}\right)
\left(\frac{v_{\rm  0}}{3 \rm km \cdot s^{-1}}\right)
\end{equation}
is a dimensionless parameter of order unity that subsumes the uncertainty in column density of $N_{\rm  D}\sim 10^{21} $ cm$^{-2}$ and $v_{\rm  0} \sim $ 3 km s$^{-1}$ is the supersonic flow speed, $f_{\rm  r}$ is the fraction of EUV photons absorbed by the recombination in the total ionization. $\Phi_{\rm  49}$ is the ionization luminosity of the star in unit of 10$^{49}$ s$^{-1}$, and $r_{\rm  d14}$ is the disc outer edge in unit of 10$^{14}$ cm.
The maximum distance can be expressed as follows:
\begin{equation} \label{6}
d_{\rm  max} \sim 10^{18} f_{\rm  FUV}^{1/2}\Phi_{\rm  49}^{1/2} \rm cm.
\end{equation}
where $f_{\rm  FUV}$ is defined as $\Phi_{\rm  FUV}=f_{\rm  FUV}\Phi_{\rm  i}$. I.e. it's a quantity to represent the fraction of the luminosity in FUV energy range and the luminosity in EUV energy range.

Here follows the developed formula about external photo-evaporation (equation \ref{2}).
In the EUV-dominated region: $d<d_{\rm  min}$ or $d>d_{\rm  max}$
\begin{equation} \label{7}
\dot{\Sigma}^{euv}(r,t)=
\begin{cases}
7\times 10^{-9} \rm M_{\rm  \odot}yr^{-1}
\left(\frac{\rm \Phi_{i}}{10^{49}}\right)^{1/2}
\left(\frac{1 \rm pc}{d}\right)\\
\left(\frac{r_{\rm  d}}{100 \rm AU} \right)^{3/2}
\frac{f_{\rm  s}}{\pi \left(r_{\rm  d}^{2}-\beta^{2} r_{\rm  g,euv}^{2} \right)}, & r>\beta \rm r_{\rm  g,euv},\\
0,& \rm otherwise.　
\end{cases}
\end{equation}
where $\Phi_{i}$ is the ionization luminosity of an external source, $d$ is the distance between external and central star, more specifically, it's a function of orbit parameters and time. We add a factor $f_{\rm  s}$, to represent the effective area fraction received the ionization luminosity, which is included the influence of inclination angle and the thickness of the disc.
\begin{equation} \label{8}
f_{\rm  s}=\max\left(\frac{h\pi r_{\rm  d}}{\pi \left(r_{\rm  d}^{2}-\beta^{2} r_{\rm   g,euv}^{2} \right)},|\cos\theta|\right)
\end{equation}
where $h$ is the thickness of the disc, and $\theta$ is the angle between the normal of the disc plane and the vector connecting central and external star. $r_{\rm  d}$ is the outer edge of gaseous disc.
\begin{equation} \label{9}
r_{\rm  g,euv}\equiv \frac {GM_{\rm  \ast}}{c_{\rm  \rm s,euv}^{2}}
\end{equation}
is the gravitational radius caused by the EUV photo-evaporation, where $c_{\rm  \rm s,euv}$ is the local sonic speed in EUV dominated region, where the gas is heated to about 10$^{4}$ K. $\beta$ is the soften factor. The first term inside bracket in equation \ref{8} represents the condition when the external star is passing through the disc plane. In this case, photons can not directly reach the surface of the disc , and most photons heat the edge of the disc. The second term represents the projected factor of disc area to receive the equivalent photo-evaporation.

In FUV dominated region, i.e. $d_{\rm  min}<d<d_{\rm  max}$, the photo-evaporation of the external star is as follows:


\begin{equation} \label{10}
\dot{\Sigma}^{fuv}(r)=
\begin{cases}
2\times 10^{-7} \rm M_{\rm  \odot} \rm yr^{-1}
\left(\frac{N_{\rm  D}}{5\times \rm 10^{21}cm^{-2}}\right )
\left(\frac{r_{\rm  d}}{100 \rm AU}\right)\\
\frac{f_{\rm  s}}
{\pi \left(r_{\rm  d}^{2}-\beta^{2} r_{\rm  g,fuv}^{2}\right )} ,&\rm r> \rm r_{\rm  g,fuv},\\
0,& \rm otherwise.
\end{cases}
\end{equation}
where $N_{\rm  D}$ is the column density and the typical value of it when dominated by FUV photons is $5-10\times 10^{21}$ cm $^{-2}$.
\begin{equation} \label{11}
r_{\rm  g,fuv}\equiv \frac {GM_{\rm  \ast}}{c_{\rm  \rm s,fuv}^{2}}
\end{equation}
is the gravitational radius caused by the FUV photo-evaporation, where $c_{\rm  \rm s,fuv}$ is the local sonic speed in FUV dominated region, where the gas is heated to about $10^{3}$ K.

From equation \ref{7} and \ref{10}, we can determine the ionization luminosity according to the distance $d$ between two stars. Obviously, $d$ varies with time and depends on the orbit parameters of flyby. As we know, the EUV photons can be easily absorbed in gas where the column density is about $10^{20}$cm$^{-2}$ \citep{2009ApJ...690.1539G}, therefore, it's hard to directly observe the EUV source and get the accurate value of $\Phi_{\rm i}$. Since $\Phi_{\rm i}$ is the most important parameters in our model, we alternatively use theoretical stellar atmosphere models to calculated the stellar spectrum. Then, we define the EUV flux at the surface of star according to \citet{2003apJ...599.1333S}, i.e.
\begin{equation} \label{12}
q_{\rm  H}=\int_{\rm  \nu_{\rm  H}}^{\nu_{\rm  m}} \frac{f_{\rm  \nu}}{h\nu} d \nu
\end{equation}
where $\nu_{\rm  H}=3.28 \times 10^{15} Hz$ is the ionization frequency for H (13.6 eV), where $\nu_{\rm  m}$ is the frequency with the energy of $100eV$. The unit of $q_{\rm  H}$ is s$^{-1}$ cm$^{-2}$,  then $q_{\rm  H}$ timing the surface area of stars is the ionization luminosity $\Phi_{\rm  i}$.
\begin{equation} \label{13}
\Phi_{\rm  i}=4\pi R^{2} q_{\rm  H}
\end{equation}
where $\Phi_{\rm  i}$ is the ionization luminosity at EUV energy range and $R$ is the radius of star. Similar to EUV flux, the ionization luminosity of FUV photons can be computed in the same way, with an energy range from 6 eV to 13.6 eV. I.e.
\begin{equation} \label{14}
q_{\rm  H\uppercase\expandafter{\romannumeral1}}= \int_{\rm  6 eV}^{\rm 13.6eV} \frac{f_{\rm  \nu}}{h\nu}d \nu
\end{equation}
\begin{equation} \label{15}
\Phi_{\rm  FUV}=4\pi R^{2} q_{\rm  H\uppercase\expandafter{\romannumeral1}}
\end{equation}
where $\Phi_{\rm  FUV}$ is the ionization luminosity at FUV energy range.
We use the Atlas9 model to calculate the stellar spectrum in both EUV and FUV bands. Atlas9 model is introduced by \citet{2004astro.ph..5087C}, and is complete for most main-sequence stars with different parameters. However, the EUV and FUV ionization luminosity for low mass stars are underestimated. E.g. for M-type stars, the ionization luminosity is much lower than the observation values. Although the realistic ionization luminosity is several orders of magnitude larger than the calculation result according to the atlas9 model, it's still not strong enough to influence the disc evolution, as we will see in section 3. Thus we mainly focus on the photo-evaporation during flybys with massive stars with $T_{\rm  eff} \geq 9500$ K.
\begin{table*}　
	\setlength{\tabcolsep}{5.5mm}
	\centering
	\begin{tabular}{lcccccr}
		\hline
		$T_{\rm  eff}$	[K]&$ \Phi_{\rm  i}$	[photons s$^{-1}$]&$\Phi_{\rm  FUV}$	[photons s$^{-1}$]&$d_{\rm  min}$	[AU]&$d_{\rm  max}$	[AU]&\emph{m}	[M$_{\rm  \odot}$]& $R$	[R$_{\rm  \odot}$]\\
		\hline
		\cline{1-7}
		38000 &2.23$\times$ 10$^{49}$ &8.25 $\times$ 10$^{49}$&1.37 $\times$ 10$^{5}$ &1.92 $\times$ 10$^{5}$ &39.4 & 17.6\\
		\cline{1-7}
		35000&8.11$\times$ 10$^{48}$ &5.26 $\times$ 10$^{49}$&8.24 $\times$ 10$^{4}$ &1.53 $\times$ 10$^{5}$ &33.3 & 15.4\\
		\cline{1-7}
		34000&5.32$\times$ 10$^{48}$ &4.45 $\times $10$^{49}$&6.68 $\times$ 10$^{4}$ &1.41 $\times$ 10$^{5}$ &31.3 & 14.7\\
		\cline{1-7}
		33000&3.29$\times$ $10^{48}$ &3.73 $\times$ 10$^{49}$&5.25 $\times$ 10$^{4}$ &1.29 $\times$ 10$^{5}$ &29.5 & 14.0\\
		\cline{1-7}
		30000&4.14$\times$ $10^{48}$ &2.02 $\times$ 10$^{49}$&2.07 $\times$ 10$^{4}$ &9.51 $\times$ 10$^{4}$ &24.2 & 12.0\\
		\cline{1-7}
		25000&1.57$\times$ $10^{46}$ &5.34 $\times$ 10$^{48}$&3.63 $\times$ 10$^{3}$ &4.89 $\times$ 10$^{4}$ &16.6 & 8.95\\
		\cline{1-7}
		19000&1.42$\times$ $10^{44}$ &6.58 $\times$ 10$^{47}$&3.45 $\times$ 10$^{2}$ &1.71 $\times$ 10$^{4}$ &9.44 & 5.76\\
		\cline{1-7}
		15000&8.72$\times$ $10^{41}$ &9.75 $\times$ 10$^{46}$&2.70 $\times$ 10$^{1}$ &6.60 $\times$ 10$^{3}$ &5.80 & 3.94\\
		\cline{1-7}
		14000&2.71$\times$ $10^{41}$ &5.04 $\times$ 10$^{46}$&1.51 $\times$ 10$^{1}$ &4.75 $\times$ 10$^{3}$ &5.03 & 3.52\\
		\cline{1-7}
		12000&2.89$\times$ $10^{39}$ &1.36 $\times$ 10$^{46}$&1.57 $\times$ 10$^{0}$ &2.47 $\times$ 10$^{3}$ &3.66 & 2.75\\		
		\cline{1-7}
		9500&4.84$\times$ $10^{36}$ &1.09 $\times$10$^{45}$&6.37 $\times$ 10$^{-2}$ &6.98 $\times$ 10$^{2}$ &2.26 & 1.89\\
		\hline
	\end{tabular}
	\caption{Parameters of external stars with different temperatures. The first column represents the effective surface temperature of the external star(T$_{\rm  \rm eff}$). The second column represents the ionization luminosity of the star in EUV range($\Phi_{\rm  i}$). The third column is the ionization luminosity of the star in FUV range($\Phi_{\rm  FUV}$). The next two columns represents  $d_{\rm   min}$ and $d_{\rm   max}$, which are the quantity to divide the EUV or FUV dominated region, according to Equation \ref{4} and \ref{6}. Note, $f_{\rm  r}=0.5$ and $f_{\rm  FUV}=\frac{\Phi_{\rm  FUV}}{\Phi_{\rm  i}}$ are adopted in this table. The last two columns represent the mass and radius of the external star, respectively, which is estimated via Stefan-Boltzmann law and Equation \ref{15} and \ref{16}. \label{Table 1}}
\end{table*}

Table \ref{Table 1} shows the parameters of different stars adopted in this paper. We use the spectral data according to atlas9 model and calculate the ionization luminosity $\Phi_{\rm  i}$ and $\Phi_{\rm  FUV}$ in EUV and FUV energy range. To estimate the mass and radius for a star with a given effective temperature $T_{\rm   eff}$, we use empirical relations in stellar astrophysics, i.e. mass-radius relation (Equation \ref{17}) and mass-luminosity relation (Equation \ref{16}). Considering Stefan-Boltzmann law $L = 4\pi \sigma R^{2} T_{\rm  eff}^{4}$, we can get the relation between $T_{\rm  eff}$, R and M approximately. As mentioned above, we don't consider low mass star, so we can make the assumption that $M>1$ M$_{\rm  \odot}$. The mass-luminosity relation from
\citet{2004adas.book.....D,2005essp.book.....S} is as follows:

%

\begin{equation} \label{16}
\begin{cases}
\frac{L}{\rm L_{\rm  \odot}} \approx 0.23 \times \left( \frac{M}{\rm M_{\rm  \odot}} \right)^{2.3}, & \emph{M}<0.43\rm M_{\rm  \odot} \\
\frac{L}{\rm L_{\rm  \odot}}= \left( \frac{M}{\rm M_{\rm  \odot}} \right)^{4}, & 0.43\rm {M}_{\rm  \odot}<\emph{M}<2\rm {M}_{\rm  \odot} \\
\frac{L}{\rm L_{\rm  \odot}}\approx 1.5  \times \left( \frac{M}{\rm M_{\rm  \odot}} \right)^{3.5}, & 2\rm M_{\rm  \odot}< \emph{M} <20 \rm {M}_{\rm  \odot} \\
\frac{L}{\rm L_{\rm  \odot}}\approx 3200  \times \left( \frac{M}{\rm M_{\rm  \odot}} \right), & \emph{M}>20\rm M_{\rm  \odot} \\
\end{cases} \\
\end{equation} 
and the mass-radius relations according to \citet{1981gask.book.....M} is as follows:
\begin{equation} \label{17}
R \sim  M^{0.7} 
\end{equation}



\section{Mass loss due to Photo-evaporation from external star}



In the case of a single flyby, we can analyze the mass loss originated by external star's photo-evaporation. Strictly, the gravitational effect between disc and external star should be considered, yet we ignored it as we set the minimum peri-center distance as 2000 AU. In such a large distance the mass loss caused by dynamical mechanisms can be neglected according to \citep{2006apJ...642.1140O,2005a&a...437..967P}, more details can be found in appendix A.

According to the Equation \ref{7} and \ref{10}, if we know the distance $d$ varying with time, the mass loss due to external star can be calculated. Integrating the total mass loss $\Delta M_{\rm  ex}$ caused by the external photo-evaporation in the duration of one flyby, we can get the relative external mass loss $\Delta M_{\rm ex, r}$. Generally, $\Delta M_{\rm ex, r}$ can be written as follows:
\begin{equation} \label{18}
\Delta M_{\rm  ex,r}= \frac{1}{M_{d0}}\int_{\rm  -t_{\rm  dur}/2}^{t_{\rm  dur}/2}\int_{\rm  r_{\rm  g}}^{r_{\rm  d}} 2 \pi r \dot{\Sigma}(r) dr dt .
\end{equation}
$M_{\rm  d0}$ is the total initial mass of gaseous disc, and is set as 0.01 M$_{\rm  \odot}$. $t_{\rm  dur}$ is the duration of a single flyby, when the photo-evaporation of external star is not ignorable. 
$r_{\rm  d}$ is the outer disc edge, and the disc is truncated when the surface density is lower than 0.01 g cm$^{-2}$.

In this section, we focus on the orbit parameters of flyby, i.e. the eccentricity $e$, inclination $inc$ and peri-center distance $q$. We will analyze the correlation between $\Delta M_{\rm ex, r}$ and these orbit parameters. We try to obtain the mass loss of gaseous disc due to external star in two cases. In the first case, only the EUV photo-evaporation works, while in another general case, both EUV and FUV contribute photo-evaporation.

\subsection{EUV photo-evaporation case}


When the external star is not hot enough, the outer boundary of FUV dominated region may be closer than the peri-center of the flyby orbit. I.e. $q > d_{\rm  max}$, and the external star only go through the EUV dominated photo-evaporation region. In this case, we can derive $\Delta M_{\rm ex, r}$ as a function of hyperbolic orbital parameters, i.e. the eccentricity $e$, inclination $inc$, and distance of peri-center $q$, we use the Kepler equation for hyperbolic orbits to derive the distance as a function of time (see appendix B). If we take $f_s=|\cos{\theta}|$ instead of equation \ref{8} for simplicity, and assume the external star arrive the peri-center at $t=0$ due to the symmetric orbit, we can get the mass loss caused by external star's EUV photo-evaporation during a flyby event:
\begin{equation} \label{19}
\Delta M_{\rm euv,r} = 
\begin{cases}
2 D\left( \frac{t_{\rm  dur}}{2}\right) /M_{d0}, & t_{\rm  dur}<2t_{\rm  c} \\
\left[ -2 D\left(\frac{t_{\rm  dur}}{2}\right)  + 4 D\left( t_{\rm  c}\right)\right]/ M_{d0}, & t_{\rm  dur} \ge 2t_{\rm  c} 
\end{cases}
\end{equation}
With the assumption of $e \gg 1$, we can derive that (see appendix B):
\begin{equation} \label{20}
\begin{aligned}
D\left(H(t)\right) \approx &C_{\rm  0}\sin(inc)\sqrt{\frac{q/1 \rm AU}{e \mu}}\left[- \frac{H}{e} + 
2 \arctan 
\left(  \tanh (H/2)\right) 
\right]
\end{aligned}
\end{equation}
where $\Delta M_{\rm euv,r}$ is the relative mass loss derived by external star's EUV photons. $H(t)$ is the hyperbolic eccentric anomaly as a function of $t$( Equation \ref{Equation B10}). $t_{\rm  c}$ is the time when the external star move from the peri-center to the the disc plane($\theta=90^{\circ}$). $D(H)$ is a function of $H$, and $C_{\rm  0}=2.3 \times 10^{-4}$  M$_{\rm  \odot}$ $(\frac{\Phi_{i}}{10^{49}})^{1/2}(\frac{r_{\rm  d}}{100 \rm AU})^{3/2}$. $q$ in unit of AU is the peri-center distance between central and external star. Dimensionless parameter $\mu$  is the total mass of the two stars, in unit of M$_{\odot}$.The mass loss is not convergent if $H \to \infty$, because in Equation \ref{7} the mass loss rate derived by external EUV photo-evaporation $\dot{M}_{\rm ex,euv} \propto d^{-1}$ \citep{1998apJ...499..758J}. However, the increase of H is a logarithms function of time $t$ (see \ref{Equation B10}). $H$ will increase slowly when the external star is very far away from the disc. Furthermore, considering that the lifetime of the disc is finite, the time $t$ as well as $H$ have an upper limits rather than infinity. 
\begin{figure*} 
	\centering
	\includegraphics[width=0.9\linewidth]{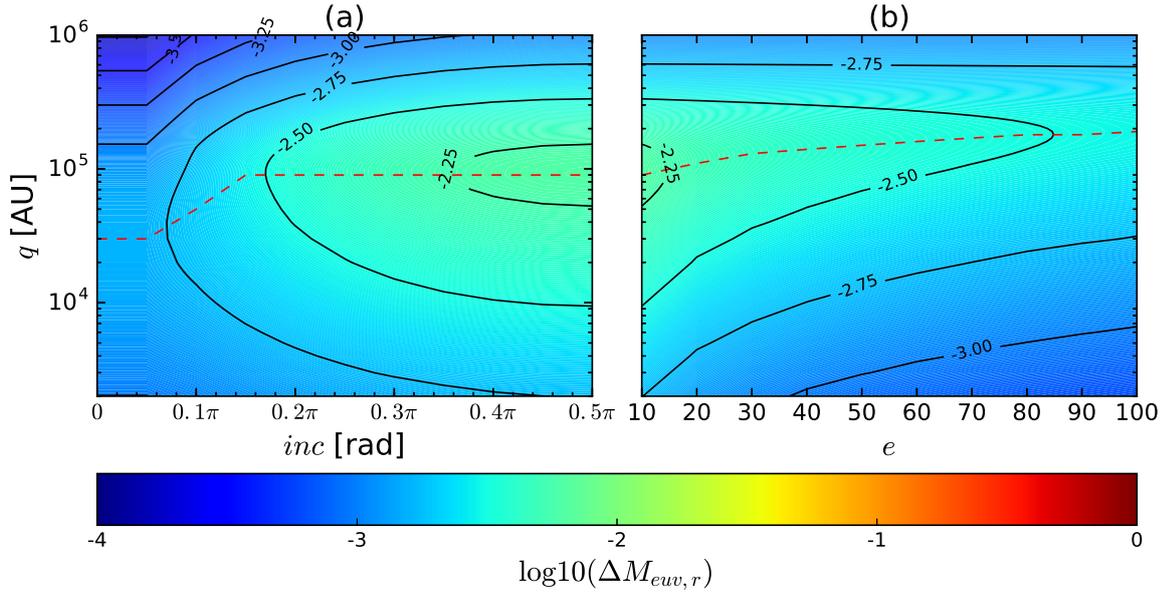}
	\caption{The correlation between  $\Delta M_{\rm  euv,r}$ (relative mass loss rate caused by external EUV photo-evaporation) and orbit parameters.  The temperature of external star is set as 19000 K, i.e. $\Phi=1.42\times10^{44}$ photons s$^{-1}$. In panel (a), using Equation \ref{20} with a fixed $e=10$, we shows how $\Delta M_{\rm  euv,r}$ varies with $q$ and $inc$. The red dashed line represents the correlation between $q$ and $inc$ when $\Delta M_{\rm  euv,r}$ achieves maximum values. In panel (b), fixing $inc=0.5\pi$, we shows how $\Delta M_{\rm  euv,r}$ varies with $q$ and $e$. The red dashed line represents the correlation between $q$ and $e$ when $\Delta M_{\rm  euv,r}$ achieves maximum values. \label{Figure 2} }
\end{figure*}

\begin{figure*} 
	\centering
	\includegraphics[width=0.9\linewidth]{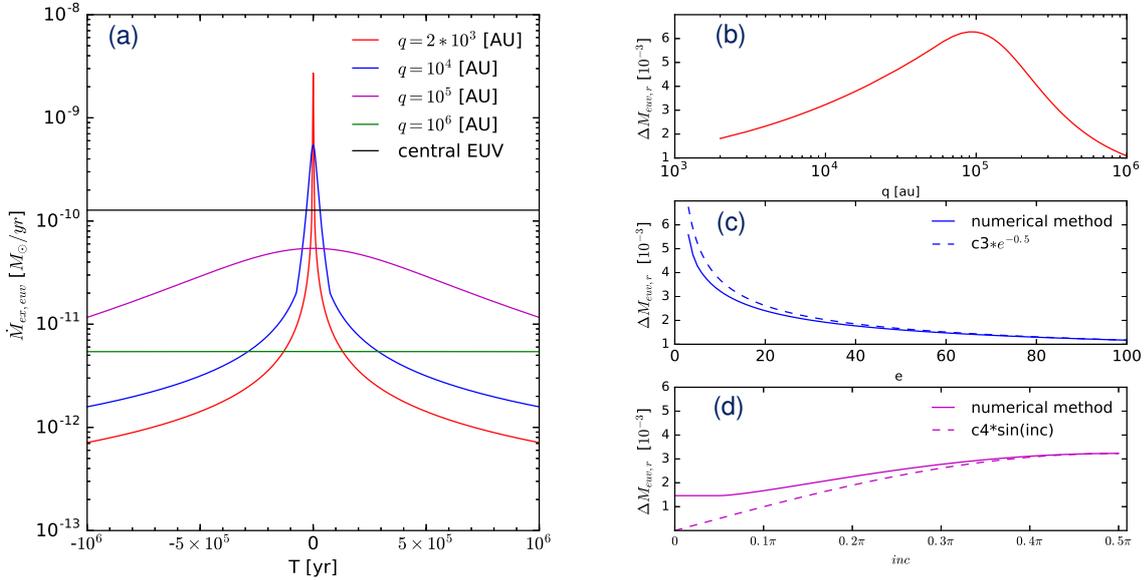}
	\caption{The correlation between  $\Delta M_{\rm  euv,r}$ (relative mass loss rate caused by external EUV photo-evaporation) and orbit parameters : $q$, $e$ and  $inc$ (direct figure). The temperature of external star is set as 19000 K. In plane (a), the disc mass loss rate caused by the external stars changes with time. With the same $inc=0.5\pi$ and $e=10$, smaller $q$ lead to larger mass loss around peri-center, however the duration around peri-center is much shorter. The black line is the mass loss rate of center star, assuming the ionization luminosity of center star is $10^{41} $ photons s$^{-1}$. In panels (b), (c) and (d), solid lines represent the relation between $\Delta M_{\rm  euv,r}$ and $q$, $e$ and $inc$ according to numerical method, respectively. And in panel (c) and (d), the dashed line shows the simple approximation of the equation \ref{22}, constant $c_{3}$, and $c_{4}$ is according to the numerical value $\Delta M_{euv,r}(inc=0.5\pi)$ and  $\Delta M_{euv,r}(e=100)$.   \label{Figure 3}} 
\end{figure*}
According to the equation \ref{7}, \ref{8} and \ref{19}, we plot Figure \ref{Figure 2} via numerical calculation, which describes the correlations between $\Delta M_{\rm  euv,r}$ and orbital parameters $q$, $inc$, $e$. Other parameters are fixed in simulations, i.e. the initial disc mass $M_{\rm  d0}=0.01$ M$_{\rm  \odot}$, with an outer boundary of 100 AU. The eccentricity of hyperbolic eccentricity $e=10$. The duration of a single flyby is fixed as $t_{\rm  dur}=2 $ Myr. 

In the Figure \ref{Figure 2}, adopting $\Phi_{i}=1.42\times10^{44}$ photons s$^{-1}$ ($T_{\rm  eff}=19000$ K), we show how the relative mass loss $\Delta M_{\rm  ex,r}=\Delta M_{\rm  euv,r}$ varies with $inc$ and $q$ (panel (a)), and with $e$ and $q$ (panel (b)). The area filled with warmer colors is preferred to generate higher mass loss. For an overview, the mass loss in all cases is limited to less than few percents. Fixing the pericenter $q$, the external star moves in an orbit with higher $inc$ or lower $e$ leads to a higher $\Delta M_{\rm  euv,r}$. When the inclination is relatively small, the orbit of the flyby is nearly co-planer with disc plane, and the external photo-evaporation only influence the disc by a factor of $f_s$ (Equation \ref{8}), therefore it lead to a low total mass loss. With the increasing of eccentricity, $\Delta M_{\rm  euv,r}$ declines. With the increase of eccentricity $e$, the external star moves faster and spends shorter time around peri-center, thus the mass loss is reduced.

Note that with the increase of peri-center distance of the orbit, $\Delta M_{\rm  euv,r}$ has a maximum value when $q$ is moderate. The red dash lines in Figure \ref{Figure 2} denote the values of $q$, where the maximum mass loss is achieved. As seen in panel (a), the dash line keep horizontal except when $inc$ varies from 0.04$\pi$ to 0.15$\pi$. The range of $inc$ is the transition region of $f_{\rm  s}$ in Equation \ref{8} and influence the averaged value depending on $inc$ during a flyby events. In panel (b), with the increasing of eccentricity $e$, the peri-center of maximum mass loss increase slightly.

Figure \ref{Figure 3} describes why there is a maximum of the total mass loss due to different $q$ during a single flyby. All the solid label lines are calculated via Equation \ref{7} and \ref{8}. In panel (a), the $inc$ and $e$ is fixed as $0.5\pi$ and 10, respectively. The stellar parameters are set the same as those in Figure \ref{Figure 2}. When the external star in orbit with smaller peri-center (e.g. 2000 AU), the mass loss rate caused by external EUV photo-evaporation $\dot{M}_{\rm  ex,euv}$ will dramatically high at peri-center, which is nearly one orders of magnitude higher than the central star. But it only lasts in a very short timescale of 0.1 Myr after the peri-center, because of the large velocity near pericenter. In other case, when $q$ is 10$^{4}$ or $10^{5}$ AU, although the maximum $\dot{M}_{\rm  euv,r}$ is  smaller than that when $q=2 \times 10^{3} $ AU, $\dot{M}_{\rm  euv,r}$ decrease much slower and lead to higher total mass loss after integration. When $q$ becomes as large as $10^{6}$ AU, the mass loss rate is too small, even smaller than one tenth of the mass loss due to the central star. thus the total mass loss is pretty low,  although the mass loss rate decays slowly for several Myr. As shown in panel (b) more clearly, the relative mass loss becomes maximum at $q\sim 10^{5}$ AU. In the panels (c) and (d) in Figure \ref{Figure 3}, the solid lines show the correlation between relative mass loss rate $\Delta M_{\rm  euv,r}$ and parameters $e$, $inc$, respectively. Other orbital parameters are fixed as shown in these panels. The correlations in panels (b) and (c) can be obtained from the analysis results in Equation \ref{20} and Equation \ref{Equation B10}. I.e. with the assumption of e$\gg$1, if $\frac{H}{e}\ll \pi/2$, equation \ref{Equation B12} becomes $\Delta {M}_{\rm  euv,r} \propto e^{-0.5}\sin{(inc)}$. Obviously, the theoretical dashed lines is very close to the solid lines in panel (c) and (d), when $e$ and $inc$ is large.  Therefore, the analysis results in Equation \ref{20} is valid to explain the influences on mass loss due to different orbital parameters.


\subsection{Both EUV and FUV photo-evaporation case}

\begin{figure*} 
	\centering
	\includegraphics[width=0.9\linewidth]{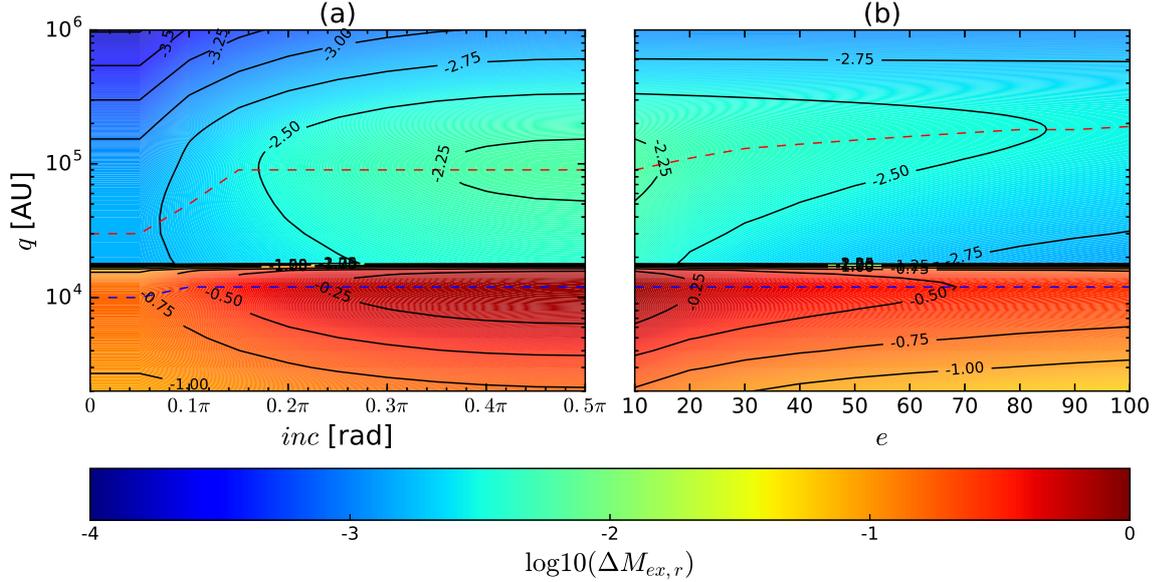}
	\caption{The same setting with Figure \ref{Figure 2}, the correlation between $\Delta M_{\rm  ex,r}$ relative mass loss, which caused by both EUV and FUV photo-evaporation, and orbit parameters : $q$, $e$ and  $inc$. In both panels, when $q > 17000 $ AU, only the EUV photo-evaporation dominates the mass loss. While $q<17000 $ AU, the FUV dominates the significant mass loss. The transition region is very narrow. \label{Figure 4}} 
\end{figure*}

During a single flyby, the external hot star with larger $d_{\rm  max}$ probably pass through both EUV and FUV regions, so there are two critical time $t_{\rm  1}$ and $t_{\rm  2}$ ($t_1<t_2$) when the external star arrived at the boundary between the EUV and the FUV dominated region. Note we set the time $t=0$ when the external star is at peri-center, and $t_{\rm  1}=0$ if $q>d_{\rm  min}$. $t_{\rm  1}$ and $t_{\rm  2}$ is tightly related to the $d_{\rm  min} $ and $d_{\rm  max}$ mentioned in Table \ref{Table 1}. According to Equation \ref{19}, the relative EUV dominated mass loss is $\Delta M_{\rm  euv,r}[H(t_{\rm  dur/2})]-\Delta M_{\rm  euv,r}[H(t_{\rm  2})]+\Delta M_{\rm  euv,r}[H(t_{\rm  1})]$, if $t<t_{1}$ and $t>t_{2}$. And similar with Equation \ref{19}, if FUV works during the whole flyby events, the relative mass loss originated by the FUV photo-evaporation can be derived in the similar way to the Equation \ref{19}:

\begin{equation} \label{21}
\Delta M_{\rm  fuv,r} =
\begin{cases}
2  E\left( t\right) /M_{d0}, & t<t_{\rm  c} \\
\left[-2   E\left(t\right)  + 4 E\left( t_{\rm  c}\right)\right]/M_{d0}, & t \ge t_{\rm  c} 
\end{cases} 
\end{equation}
and
\begin{equation} \label{22}
E\left(t \right) = C_{\rm  1}\sin(inc)\sqrt{\frac{(q/1\rm AU)^{3}}{(e-1)^{3}\mu}}\left[eH-\sinh(H)\right],
\end{equation}
where $t$ is the time after the star arrives at the pericenter. $C_{\rm  1}=3.2\times 10^{-8}$ M$_{\rm  \odot}$yr$^{-1}$$(\frac{N_{\rm  D}}{5\times 10^{21}\rm cm^{-2}} )(\frac{r_{\rm  d}}{100 \rm AU})$. We only considered $f_s$ as $\left|\cos(\theta)\right|$ for simplicity instead of the Equation \ref{8}. The relative mass loss originated in FUV photo-evaporation dominated region can be expressed as $\Delta M_{\rm  fuv,r}(t_{\rm  2})-\Delta M_{\rm  fuv,r}(t_{\rm  1})$. Then we combine these two parts of the mass loss, and finally get the analytical form of $\Delta M_{\rm ex,  r}$ for both EUV and FUV photo-evaporation:

\begin{equation} \label{23}
\Delta M_{\rm ex, r}=\Delta M_{\rm  euv,r}\left(t_{dur}/2\right)-\Delta M_{\rm  euv,r}\left(t_{\rm  2}\right)+\Delta M_{\rm  euv,r}\left(t_{\rm  1}\right)+\Delta M_{\rm  fuv,r}\left(t_{\rm  2}\right)-\Delta M_{\rm  fuv,r}\left(t_{\rm  1}\right)
\end{equation}
The analytical formula of $\Delta M_{\rm ex, r}$ is complicated. Directly solving the Kepler function is a better way. We use numerical method to solve the Kepler equation, and calculate the total mass loss during a flyby event via Equation \ref{7} and \ref{10} alternatively. The calculation results are shown in Figure \ref{Figure 4}. There is an obvious boundary when $q\sim 1.7\times 10^{4}$ AU, where is the boundary $d_{\rm  max}$ between FUV and EUV dominated regions for stars with $T_{\rm  eff}=19000$ K. Above $q\sim 1.7\times 10^{4}$ AU, only EUV photo-evaporation works and it's similar with Figure \ref{Figure 2}. The mass loss during a single flyby is only few percents. Under $q\sim 1.7\times 10^{4}$ AU, the FUV dominated the mass loss. Obviously, the mass loss is extremely large and even close to the initial disc mass, during single flyby events with small $e$ or large $inc$, when $q$ is around $10^{4}$ AU. Therefore, if FUV photo-evaporation works, the disc will disperse in a shorter timescale. Additionally, similar to Figure \ref{Figure 2}, there is a moderate $q\sim 10^{4}$ AU when mass loss rate becomes maximum. The dependence of the moderate values on parameters $inc$ or $e$ is resemble to that in EUV dominated region. Note that we fix the outer edge of disc $r_d=100$ AU to estimate the photo-evaporation. However, $r_d$ may be reduced to less than $100 $ AU during flybys, because of the large photo-evaporation of FUV (as shown in section 4), and lead to smaller mass loss than that shown in Figure \ref{Figure 4}.

In most cases, the photo-evaporation rate of FUV dominates the total mass loss, thus Equation \ref{23} can be simplified as follows:

\begin{equation} \label{24}
\Delta M_{\rm ex, r} \approx \Delta M_{\rm  fuv,r}\left(t_{\rm  2}\right)-\Delta M_{\rm  fuv,r}\left(t_{\rm  1}\right).
\end{equation}

Furthermore, we can conclude that the total mass loss during a flyby events mainly depend on the duration when the FUV photo-evaporation and the averaged factor of $f_{\rm  s}$, if the outer disc boundary keeps constant. 

\begin{table}
	\begin{center}
		\caption{Input parameters for central star and initial disc}\centering
		\begin{tabular}{p{3.5cm}<{\centering}p{3.5cm}<{\centering}}
			\hline
			\hline
			Stellar mass & 0.5 M$_{\odot}$ \\
			Ionization luminosity & 10$^{41}$ s$^{-1}$ \\
			disc mass  & 0.01 M$_{\odot}$ \\ 
			Surface density & $\Sigma(r) \propto r^{-3/2}$ \\
			Inner disc boundary & 0.1 AU \\
			Outer disc boundary & 100 AU \\
			\hline
			\hline
		\end{tabular}
	\end{center}
\end{table}
\section{Disc evolution under photoevaporations}
In section 3, we analyze the mass loss due to the external flyby star. However, disc evolution are correlated with the photo-evaporation from the central star and the viscosity diffusion in the gaseous disc. In this section, we combine the viscosity and photo-evaporation from both central and external stars, and obtain the disc evolution via numerical simulations. We will investigate how the ionization luminosity of external star and the hyperbolic orbit parameters influence the disc evolution, especially the disc lifetime and the timescale for gap-opening, which are important for planet formations.

In the one dimension gaseous disc model introduced in Equation \ref{2}, we set the central star mass $m_{\rm  c}=0.5$ M$_{\rm  \odot}$ \citep{2005MNRaS.358..742S}. The ionization luminosity of the central star $\Phi_{\rm  c}=10^{41}$ photons s$^{-1}$, and is assumed to be constant. It's a typical value used in many central EUV photo-evaporation models \citep{2001MNRAS.328..485C}. The initial disc mass $M_{\rm  d0}=0.01$ M$_{\rm  \odot}$ which is typical according to \citet{2001MNRAS.328..485C}. And the gaseous disc is in the range of [0.1,100] AU. In the alpha-disc model, the dimensionless viscous parameter $\alpha$ is set as 0.001. 

The initial disc surface density distribution obeys the power law with an index of $-1.5$ \citep{2011aRa&a..49...67W} which considered as the initial surface density distribution of the solar
system, i.e. $\Sigma\left( r\right) = \Sigma_{\rm  0} （\frac{r}{1 \rm AU}）^{-3/2}$.  
Constant $\Sigma_{\rm  0}$=730 g cm$^{-2}$ to ensure the initial disc mass is 0.01 M$_{\rm  \odot}$. 

\begin{figure} 
	\centering
	\includegraphics[width=0.5\linewidth]{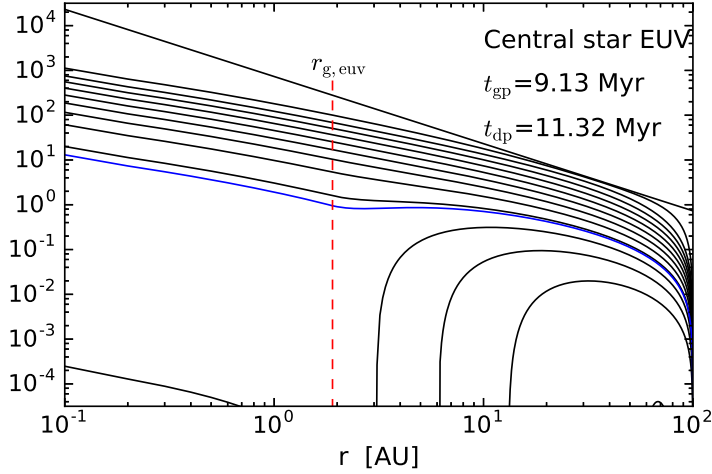}
	\caption{The snapshot of the disc evolution only driven by the central star's photo-evaporation and viscosity. The solid black lines shows the profile of surface density at 0-14 Myr from up to bottom, with equal intervals of 1 Myr, and the solid blue line shows the profile when the gap is opening. The red dashed line represents the position of the gravitational radius considering the soften factor $\beta=0.5$. Other parameters are set as follows: the initial disc mass $M_{\rm  d0}=0.01 $ M$_{\rm  \odot}$. The luminosity of central star $\Phi_{\rm  c}$=10$^{41}$ photons s$^{-1}$. In the alpha-disc, the viscous parameter $\alpha = 0.001$. \label{Figure 5}}
\end{figure}

In Figure \ref{Figure 5}, we shows a snapshot of the disc evolution only driven by the central star's photo-evaporation and viscosity. It's used to compare with other cases in this section including external photo-evaporation. In Figure \ref{Figure 5}, the gap opens at $\sim$ 2 AU after $t_{\rm  gp}=9.13$ Myr. Here, We adopt the time, when $\frac{d \log{\Sigma(t)}}{d \log{r}}$ firstly change from negative to positive value, as $t_{\rm  gp}$. After the gap-opening, the inner disc will shrink and finally disperse in short time of several $10^{5}$ yr. At the same time the gap becomes bigger and bigger. Inner boundary of the outer disc is put outward, while the outer disc edge shrink slowly as the disc is dispersing. The dispersal timescale of gaseous disc $t_{\rm  dp}$ is about 11.13 Myr, when the total outer disc mass (where the radius is larger than soften gravitational radius $\beta r_{\rm  g,fuv}\sim 36 $ AU) is less than one percent of initial outer disc mass, i.e. $<0.1$ Jupiter mass. Therefore, the gas giant formation and planet migration are hardly be influenced after $t_{\rm  dp}$. Since the gap opening occurs when the disc is nearly disperse, planet formation and migration can hardly influence by the gap-opening neither.  

In this section, we assume the flyby duration is fixed as 2 Myr(expect for the case with pericenter distance $q=10^{6}$ AU, we set the flyby duration as 10 Myr, more detailed will be found in Appendix B) and the star arrives the peri-center at 1 Myr exactly. In all the following cases, the star is as far as $> 10^{6}$ AU at both $t=$ 0 or 2 Myr, i.e. the star have passed through the FUV-dominated region and the EUV photo-evaporation effect becomes much smaller than that from center star (Figure \ref{Figure 3}).

\subsection{Internal ionization luminosity and disc viscosity}
\begin{figure} 
	\centering
	\includegraphics[width=0.5\linewidth]{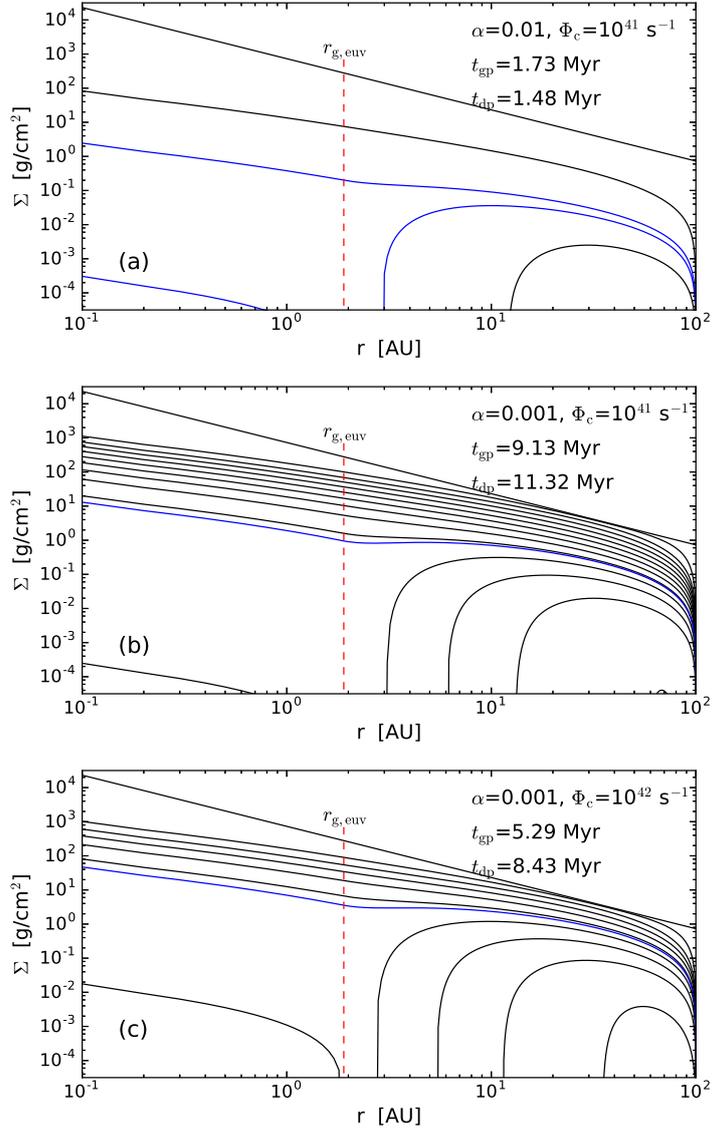}
	\caption{The disc surface density evolution with internal photo-evaporation and viscosity evolution. The red lines represent $r_{\rm  g,euv}$, the gravitational radius where gas temperature is heated to $10^{4}$ K by EUV photo-evaporation. In panel (a), ionization luminosity of central star $\Phi_{c}=10^{41}$ s$^{-1}$, dimensionless viscous coefficient $\alpha=0.01$; in panel (b), $\Phi_{\rm c}=10^{41}$ s$^{-1}$, $\alpha=0.001$; in panel (c), $\Phi_{\rm c}=10^{42}$ s$^{-1}$, $\alpha=0.001$. \label{Figure B2}} 
\end{figure}

In this subsection section, we change the dimensionless viscous coefficient $\alpha$ and ionization luminosity of central star $\Phi_{\rm c}$. Comparing panel (a) with (b) in Figure \ref{Figure B2}, $\alpha=0.01$ in panel (a) is one order magnitude larger than that in panel (b), which results in the disc dispersal time scale $t_{\rm dp}$ in panel (b) is roughly $\sim 10$ times that in panel (a). Considering the viscous evolution time scale $t_{\nu} \sim r^{2}/ \nu$, where $\nu$ is the kinematic viscosity and proportional to $r$ and $\alpha$. Thus larger $\alpha$ leads to smaller $t_{\rm \nu}$, and the disc disperse more quickly due to larger inner and outer flow in disc boundaries. In panel (a), the blue solid lines represent the surface density profiles at time $t=1.7$ Myr and 1.8 Myr, to show the outer disc shrinks quickly after gap opens. In the case with $\alpha=0.01$, $t_{\rm dp}<2$ Myr is consistent with observations that protoplanetary gas disc can disperse in 3 Myr.

In panel (c), although we increase the ionization luminosity of the central star to $10^{42}$ s$^{-1}$, it still need to wait another $\sim 2 $ Myr until gap expand to 20 AU. Thus in a disc with $\alpha=0.001$ the disc will not rapidly disperse due to a relatively slow viscous times scale. It implies the disc viscosity, instead of the internal ionization luminosity, dominates the outer disc evolution after gap opens in our model. The outer disc disperses as quickly as that in the UV-switch model, only if the disc is very viscous.

\subsection{External stellar parameters}
In this subsection we mainly concentrate on the influence of the external radiation parameters. In order to show the relation between disc evolution and radiation parameters, we fix the other orbital parameters, i.e. $inc=0.5\pi$, $e=10$ and $q=10000$ AU. 
\begin{figure*} 
	\centering
	\includegraphics[width=0.9\linewidth]{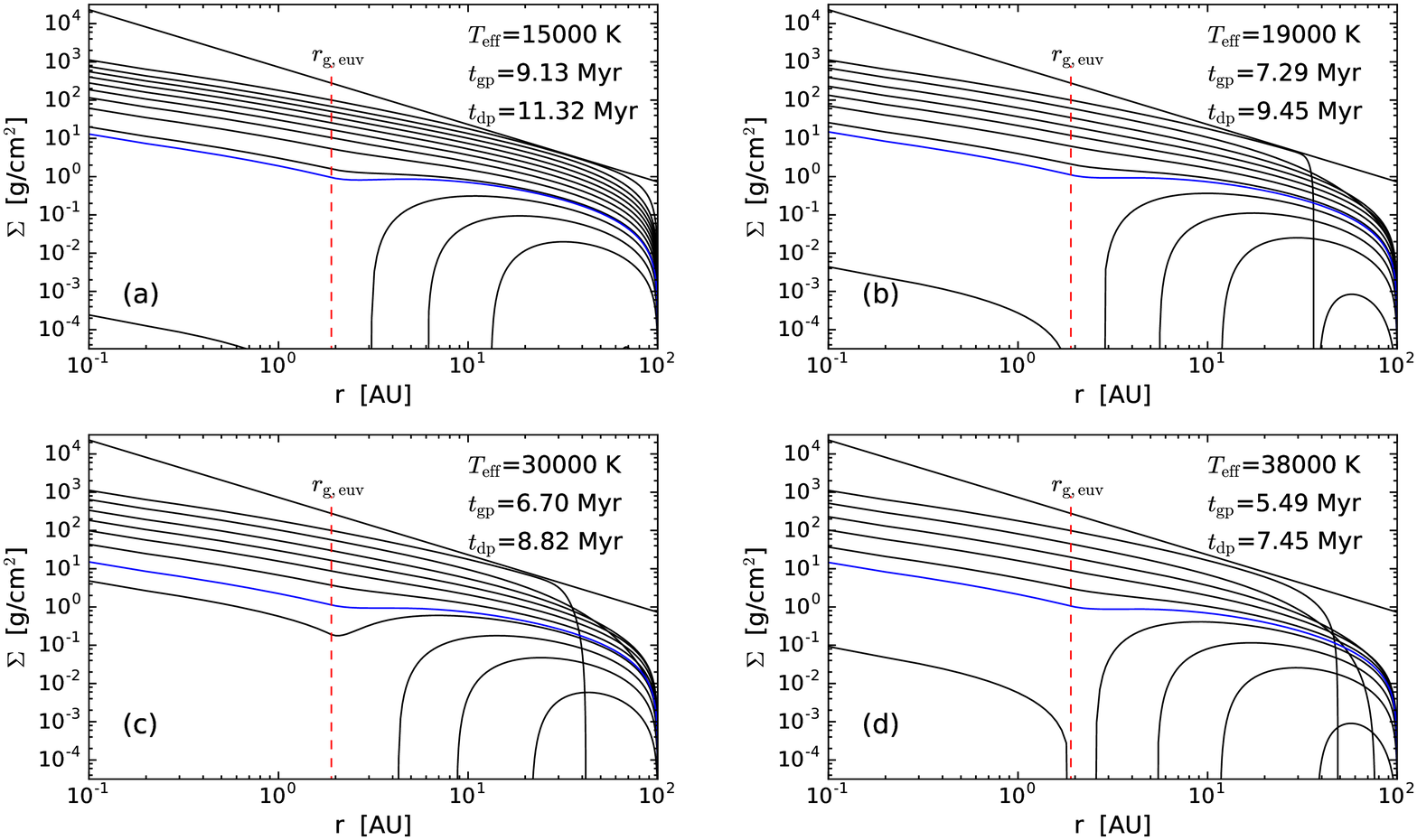}
	\caption{The disc surface density evolution with external stars with different $T_{\rm  eff}$. The red lines represent $r_{\rm  g,euv}$, the gravitational radius where gas temperature is heated to $10^{4}$ K by EUV photo-evaporation. The flyby orbit is fixed as $inc=0.5\pi$, $e=10$ and $q=10000$ AU. \label{Figure 6}} 
\end{figure*}

In figure \ref{Figure 6}, we present the disc evolution evaporated by external stars with different $T_{\rm  eff}$. For stars with temperature $T_{\rm  eff}=15000$ K, the disc evolution is the same as the standard case that only viscosity and central EUV photo-evaporation works (see Figure \ref{Figure 5}). Since the external star do not go through the FUV region, only EUV photo-evaporation works. However, the external star with $T_{\rm  eff}=15000$ K has an ionization luminosity $\sim 10^{41}$ photons s$^{-1}$, which is similar to the central star. Considering that the external star is much further than the central star, the radiation of external star can hardly change the evolution of the disc obviously. Comparing with Figure \ref{Figure 5}, we can easily see that the external stars with $T_{\rm  eff}>15000$ K shrink the outer disc edge during flyby at 1 Myr, when the star is as close as 10000 AU and accelerate the disc evolution. 

For hotter external stars, the photo-evaporation effect is much stronger, especially when the external star is close to the peri-center. Therefore, the outer disc edge shrinks rapidly in the first million year as shown in figure \ref{Figure 6}. Then compared with the three panels (b), (c) and (d), we can find that the outer disc edge in b is smaller than that in panels (c) and (d). The location of disc outer edge is determined by the balance between disc viscous diffusion and photo-evaporation of external star. For stars with $T_{\rm  eff} = 19000$ K, when the distance $d = q = 10000$ AU $\sim 1.5 \times 10^{17}$ cm) at $t = 1$ Myr, which is inside the FUV dominated region as shown in Table \ref{Table 1}. However, for stars with $T_{\rm  eff}=30000$ K or 38000 K, the distance $d = q =10000$ AU at time 1 Myr is in EUV dominated region. Note the mass loss rate generated by the FUV photons has one magnitude larger than EUV photons of the hottest stars with $T_{\rm  eff}=38000$ K (see Equation \ref{7} and \ref{10}, $d$=0.5 pc). Thus the outer edge becomes smaller for external stars of 19000 K than other stars at $t=1$ Myr. The disc depletes around 7.45 Myr for the extreme case of $T_{\rm  eff}=38000$ K, but it's still typical dispersal timescale for proto-planetary disc.

For the overview of the figure \ref{Figure 6}, massive stars with higher ionization luminosity will generate much more mass loss, and accelerate the disc evolution. Meanwhile, the time scale of gap opening is put forward by external star with higher $T_{\rm  eff}$.

\begin{table} 
	\begin{center}
		\begin{tabular}{cccc}
			\hline
			$T_{\rm   eff}$ [K] & $t_{\rm  dp}$ [Myr] & $t_{\rm  gp}$ [Myr] & $\tau_{\rm  fuv}$ [Myr]\\
			\hline
			\cline{1-4}
			9500&11.3&9.13&0 \\
			\cline{1-4}
			15000&11.3&9.13&0 \\
			\cline{1-4}
			19000&9.45&7.29&0.0444 \\
			\cline{1-4}
			25000&9.21&7.06&0.1184 \\
			\cline{1-4}
			26000&9.12&6.97&0.1224 \\
			\cline{1-4}
			29000&8.87&6.74&0.1392 \\
			\cline{1-4}
			30000&8.83&6.70&0.1596 \\
			\cline{1-4}
			33000&8.33&6.24&0.1474 \\
			\cline{1-4}
			34000&8.14&6.07&0.1386 \\
			\cline{1-4}
			35000&8.01&5.96&0.1284 \\
			\cline{1-4}
			38000&7.45&5.49&0.0922 \\    
			\hline
		\end{tabular}
		\caption{Timescales due to single flyby with different stars. The first column is the effective surface temperature. The second column is the dispersal timescale of the disc, I.e the time when the disc has lost the 99 percent of it's initial mass outside 36 AU. The third column is the timescale of when the gap forms. Last column $\tau_{\rm  fuv}$ represents the duration when the external star go through the FUV photo-evaporation dominated region. \label{Table 2}}
	\end{center}
\end{table}



As we mentioned above, the dispersal of the disc is mainly driven by the central star's photo-evaporation, external star's photo-evaporation and the evolution of the viscosity. The combination of the central star's photo-evaporation and viscosity lead to open a gap near the soften gravitational radius $r_{\rm  g,euv}$. Considering the influence of external star's photo-evaporation, different disc evolution may lead to different gap-opening time $t_{\rm  gp}$. 

In table \ref{Table 2}, we list the dispersal time scale $t_{\rm  dp}$ and the gap-opening time $t_{\rm  gp}$ in different single flyby cases, based on numerical simulations. $t_{\rm  dp}$ is obtained when the total disc mass is less than 30 Earth mass, i.e. one percentage of the initial disc mass. After $t_{\rm  dp}$, there is not enough gas to form giants planets. From Table \ref{Table 2},  $t_{\rm  gp}$ is alway smaller than $t_{\rm  dp}$. After the gap-opening, the planets formed in such disc will undergo different migrations before the gas disperses. Therefore, the gap can affect the planet architectures. With the increasing of the $T_{\rm  eff}$ of the flyby star, the time scale of gap open $t_{\rm  gp}$ declines. However, $t_{\rm  gp}$ is close to the dispersal timescale, i.e. gap opens when the disc mass is much less than the initial mass. It implies that only if the planet forms outside the photoevaporative gap and not long before the gap forms, it could expected the gap will effectively halt the planet's migration once the planet and gap near one another. Considering that gaps in our simulations form fairly late in the disc lifetime and it's larger than the typical giant planet forming time scale. Thus we indicate that gap only influences the formation and orbital architecture of planets limitedly.

As shown in Equation \ref{7} and \ref{10}, the photo-evaporation via FUV is much larger than that via EUV, and the mass loss rate is independent on distance and luminosity of external star. Thus the total mass loss mainly depends on how long the external star go through the FUV dominated regions (denoted as $\tau_{\rm  fuv}$). We calculate $\tau_{\rm  fuv}$ for different cases in Table \ref{Table 2}, and see a negative correlation between $\tau_{\rm  fuv}$ and $t_{\rm  dp}$ when $T_{\rm  eff}<30000 $ K. For the cases of $T_{\rm  eff}=9500 $ K and 15000 K, where $\tau_{\rm  fuv}=0$, only the EUV photo-evaporation works, the $t_{\rm  dp}$ is the same with the case in Figure \ref{Figure 5}. I.e. the duration of photo-evaporation via external FUV determined the disc dispersal timescale. The positive correlation between $\tau_{\rm  fuv}$ and $t_{\rm  dp}$ when $T_{\rm  eff}>30000 $ K is because the mass loss due to EUV becomes more and more important.

Why the gap opens is because the mass transfer rate due to viscosity is less than the mass loss rate due to photo-evaporation of central star at certain radius after the flyby event. The mass loss originated by the central star's photo-evaporation is proportional to r$^{-2.5}$ (see Equation \ref{3}). During disc dispersion, the surface density of the disc declines, and the mass transfer rate caused by viscosity decreases at the same time. However the mass loss caused by central star $\dot{M}_{\rm  c}$ is nearly constant, because we set the ionization luminosity of central star $\Phi_{\rm  c}$ as constant. When $\dot{M}_{\rm  c}=\dot{M}_{\rm  v}$, where $\dot{M}_{\rm  v}$ is the mass transfer rate caused by viscosity (i.e. gas accretion rate from the central star), the gap begins to open. 
\begin{figure} 
	\centering
	\includegraphics[width=0.5\linewidth]{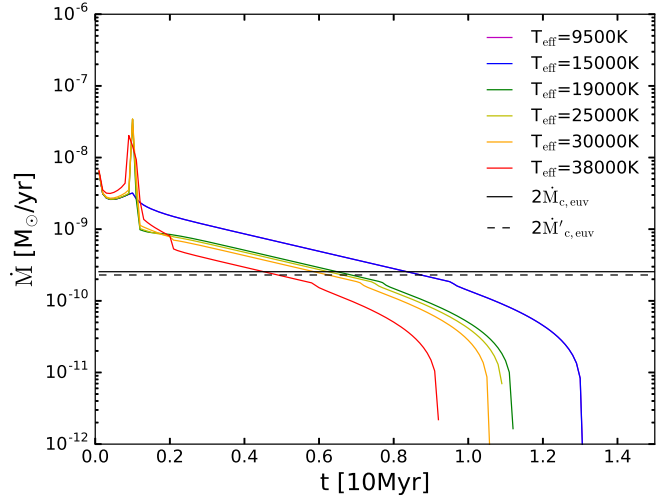}
	\caption{The correlation between the total mass loss rate of the disc($\dot{M}$), with time in different external radiation field. Parameter settings : $e$=10, $q$=10000 AU and $inc$=0.5$\rm \pi$. The lines in magenta, blue, green, yellow, orange and red represent the effective surface temperature of 9500 K, 15000 K, 19000 K, 25000 K, 30000 K and 38000 K, respectively. The black solid line and dashed line all represents the mass loss rate generated by the central star's EUV photo-evaporation. The difference between the two is that the dashed line considering the change of the disc edge times an effective fraction 0.9, while the solid line doesn't. \label{Figure 7}}	
\end{figure}
In Figure \ref{Figure 7}, we show the correlation between the total disc mass loss rate $\dot{ M}$. In the first two million years, $\dot{ M}$ includes mass loss of accretion, internal photo-evaporation and external photo-evaporation; after 2 Myr, when flyby events ends, the $\dot{ M}$ equals to the mass loss driven by accretion and internal photo-evaporation. The total mass loss rate varies with time $t$ for different external stars. The peaks of $\dot{ M}$ inside 2 Myr, is because of photo-evaporation from external star. To estimate the time when gap-opening occurs in the disc, we adopted a criterion as follows:

\begin{equation} \label{25}
\dot{\rm M}=2\dot{\rm M}_{\rm  c,euv}
\end{equation}

Note the total mass loss rate includes both mass accretion due to viscosity $\dot{M}_{\rm  v}$ and mass loss due to photo-evaporation of central star $\dot{M}_{\rm  c,euv}$, after flyby events. The Equation \ref{25} means $\dot{M}_{\rm  c,euv}=\dot{M}_{\rm  v}$. We plot the criterion Equation \ref{25} (estimated via Equation \ref{3}) as horizontal solid line in Figure \ref{Figure 7}, which intersects with different lines of $\dot{M}$ due to different flyby stars, the crossing points represent when the gap is opening. The criterion gives $t_{\rm  gp}$ in Figure \ref{Figure 6}, which are consistent with the simulation results of $t_{\rm  gp}$ in Table \ref{Table 2}.

After the gap opening, there will be a sharp decreasing of $\dot{M}$ in every case. It is because the inner disc can no longer get the mass from outer disc after gap-opening, and the inner disc is accreted onto the central star in a short timescale. The inner disc will disperse quickly and the disc lifetime is determined by the outer disc evolution.

\subsection{Orbit parameters}
We will discuss the influence on disc evolution due to orbit parameters, including eccentricity $e$, inclination between orbit plane of external star and the disc plane $inc$ and distance of peri-center $q$. In this subsection, the effective temperature of external stars is fixed as 19000 K in the simulations.
\subsubsection{eccentricity}
\begin{figure*} 
	\centering
	\includegraphics[width=0.9\linewidth]{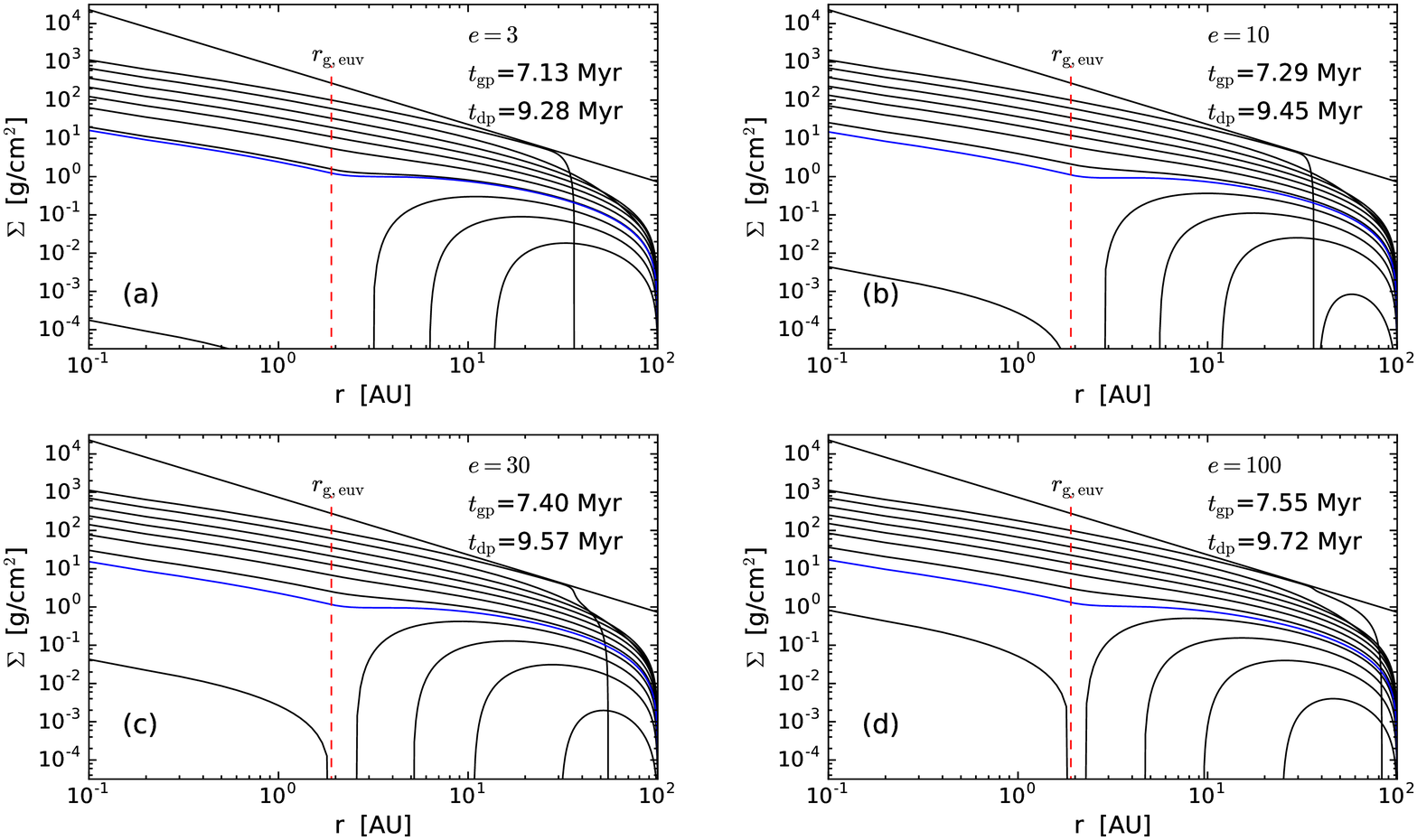}
	\caption{Similar with Figure \ref{Figure 6},the snapshots of the surface density are plotted every 1 Myr, from top to bottom. The eccentricity are 3, 10, 30, and 100 in the four panels. We set $inc=0.5\pi$, $q=10000$ AU and $T_{\rm  eff}=19000$ K in the simulations.\label{Figure 8}} 
\end{figure*}

In this subsection, we fix the inclination as $0.5\pi$ and set the pericenter $q=10000 $ AU. As shown in figure \ref{Figure 8}, in the first million year when the flyby comes to the peri-center, the outer disc edge shrinks with the decrease of the eccentricity i.e. higher eccentricity $e \ge 30$, the larger outer disc edge $r_{\rm  d}$ at the time of 1 Myr. While when $e=$3 and 10 the outer disc edge remain nearly the same around 36 AU(Which is near the soften gravitational radius $r_{\rm  g,fuv}$ where gas temperature is heated to 1000K). Furthermore, different $e$ lead to similar time scales of gap opening $t_{\rm  gp}$, which is about 2 Myr earlier than the model without external star in Figure \ref{Figure 5}. The disc dispersal time scale $t_{\rm  dp}$ is also similar, and is about 1 Myr earlier than that in Figure \ref{Figure 5}. 

Considering that the effective temperature of the external star $T_{\rm  eff}=19000$ K, the FUV dominated region is close to pericenter, therefore both EUV and FUV photo-evaporation should be included. According to analytical results in Figure \ref{Figure 4} with the same external stellar parameters, the larger $e$ is, the smaller $\Delta M_{\rm  ex}$ is. The maximum $\Delta M_{\rm  ex}$ is $\sim$ 14 per cent of total disc mass in the case of $e=3$. If $e=100$, $\Delta M_{\rm  ex}$ becomes only 2.5 of total disc mass. After the flyby events (or after 2 Myr), the viscosity and central star dominated the evolution of the disc. The disc density profiles become more and more homogeneous due to viscosity evolution, and finally the disc have similar $t_{\rm  dp}$ and $t_{\rm  gp}$.

To explain the outer boundary of gaseous disc at 1 Myr, if we assume the viscosity and photo-evaporation is balance at the disc edge, we can predict that the disc outer edge should be the same, because the same viscosity and external star radiation at 1 Myr. It's consistent with $e=3$ and 10, where the outer disc edges are both around 36 AU. However, when $e=30$ and 100, the disc outer edge becomes approximate 54 and 80 AU, respectively. For the cases of large eccentricities, the star goes through the FUV dominated region too fast, and there is not enough time to clear the mass outside 36 AU, thus the edge becomes larger as the eccentricity increases. Here we do a rough calculation of the mass loss during the star go through the FUV dominated region (from 10000 to 17000 AU according to Table \ref{1}). when the star go into the FUV region, i.e. 17000 AU, it will take 8600 years to arrive at the pericenter when $e=30$ (see appendix B, Equation \ref{Equation B3} and \ref{Equation B10}). Since the mass loss rate of star is constant, i.e. $\dot{M}= 10^{-9} \frac{r_{\rm  d}}{1 \rm AU} $ M$_{\rm  \odot}$ \rm yr$^{-1}$ according to Equation \ref{10}, we adopted a maximum $r_{\rm  d}=100 $ AU. The total mass loss until 1 Myr is 8.6 $\times 10^{-4}$  M$_{\rm  \odot}$. The disc mass outside 36 AU should be 0.004 M$_{\rm  \odot}$ , which is much larger than the total mass loss, thus the external star only truncated the disc at a larger distance. The same reason for $e=100$ case, where the external star moves faster, and take only 1200 year to stay inside the FUV dominated region in the first 1 Myr. Therefor very few mass loss in the first 1 Myr, and the disc outer edge is around 80 AU.

To summary, Both the analysis in section 3 and the numerical simulation in this section lead to the result that external star with higher eccentricity, gives few changes during disc evolution. However, the influence on both gap-opening time and disc dispersal time due to eccentricity is very limited. Combine with Figure \ref{Figure 6}, even for external star with $T_{\rm  eff}=38000 $ K with $e=10$, the disc dispersal time $t_{\rm  dp}$ changes a little comparing with other external stars. Thus we can conclude the disc dispersal time is hardly changed after a single flyby events with pericenter $q=10000 $ AU, no matter how hot the external star ($T_{\rm  eff}\le38000 $ K) is or how eccentric the orbit is ($e\ge3$).

\subsubsection{inclination}
\begin{figure} 
	\centering
	\includegraphics[width=0.5\linewidth]{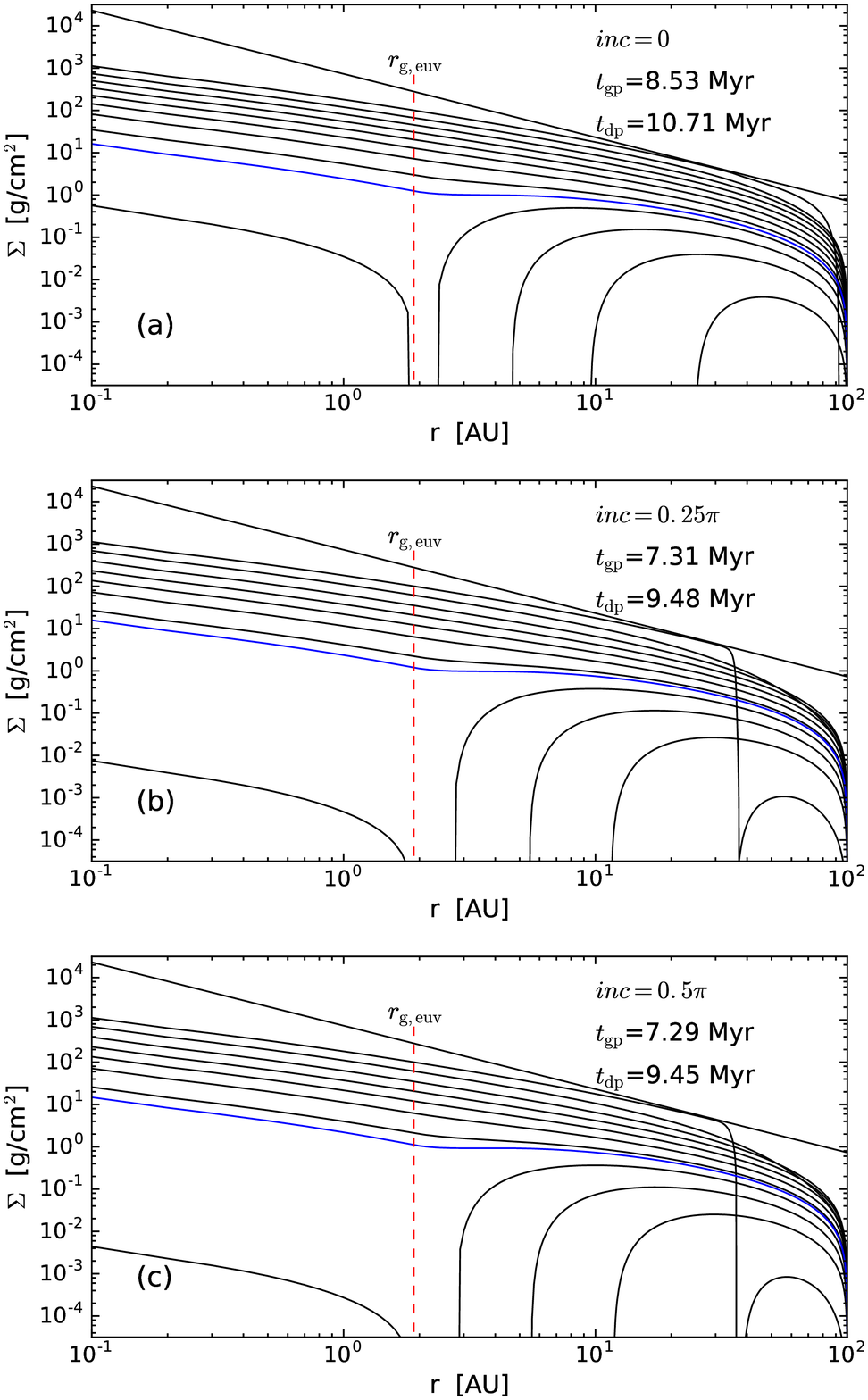}
	\caption{Similar with Figure \ref{Figure 8}, the snapshots of the surface density are plotted every 1 Myr, from top to bottom. The inclination angle is set as 0$\pi$, 0.25$\pi$, 0.5$\pi$, while $e$ is fixed as 10 and $q$ is fixed as 10000 AU. The external star has a $T_{\rm  \rm eff}$ of 19000 K. \label{Figure 9}} 
\end{figure}

The inclination we define here is the angle between the orbital plane of external star and the disc plane. The eccentricity is set as 10, while $q=10000 $ AU. In figure \ref{Figure 9}, the increase of the inclination angle will accelerate the disc evolution. With higher inclination angle $inc$, the mass loss caused by external star will increase according to the factor $\sin{(inc)}$ in Equation \ref{20} and \ref{22}. Gap opens at earlier time, and consequently the time scale of disc dispersal $t_{\rm  dp}$ decreases. In the cases of $inc=0.5\pi$ and $inc=0.25\pi$, both $t_{\rm  dp}$ and $t_{\rm  gp}$ are nearly the same, while in the case that $inc=0$, $t_{\rm  dp}$ and $t_{\rm  gp}$ is about only 1 Myr later. 

The dependence on $inc$ are consistent with the analysis results in Figure \ref{Figure 4} panel (a). According to equation \ref{20}, $\sin{0.25\pi} \sim \sin{0.5\pi} = 1$, while when $inc=0$ (according to Equation \ref{8}), $f_{\rm  s} \sim 0.05$(assume $h \sim 0.05 r_{\rm  d}^{1.25}$), therefore in cases of $inc=0.5\pi$ and $inc=0.25\pi$, the mass loss caused by external photo-evaporation $M_{\rm  ex}$ is one magnitude order larger than that in the case of $inc=0$. Even if $inc=0.5\pi$, the $t_{\rm  dp}$ as long as $\sim 10 $Myr can not influence the planet formation obviously. 

\subsubsection{peri-center distance}
\begin{figure*} 
	\centering
	\includegraphics[width=0.9\linewidth]{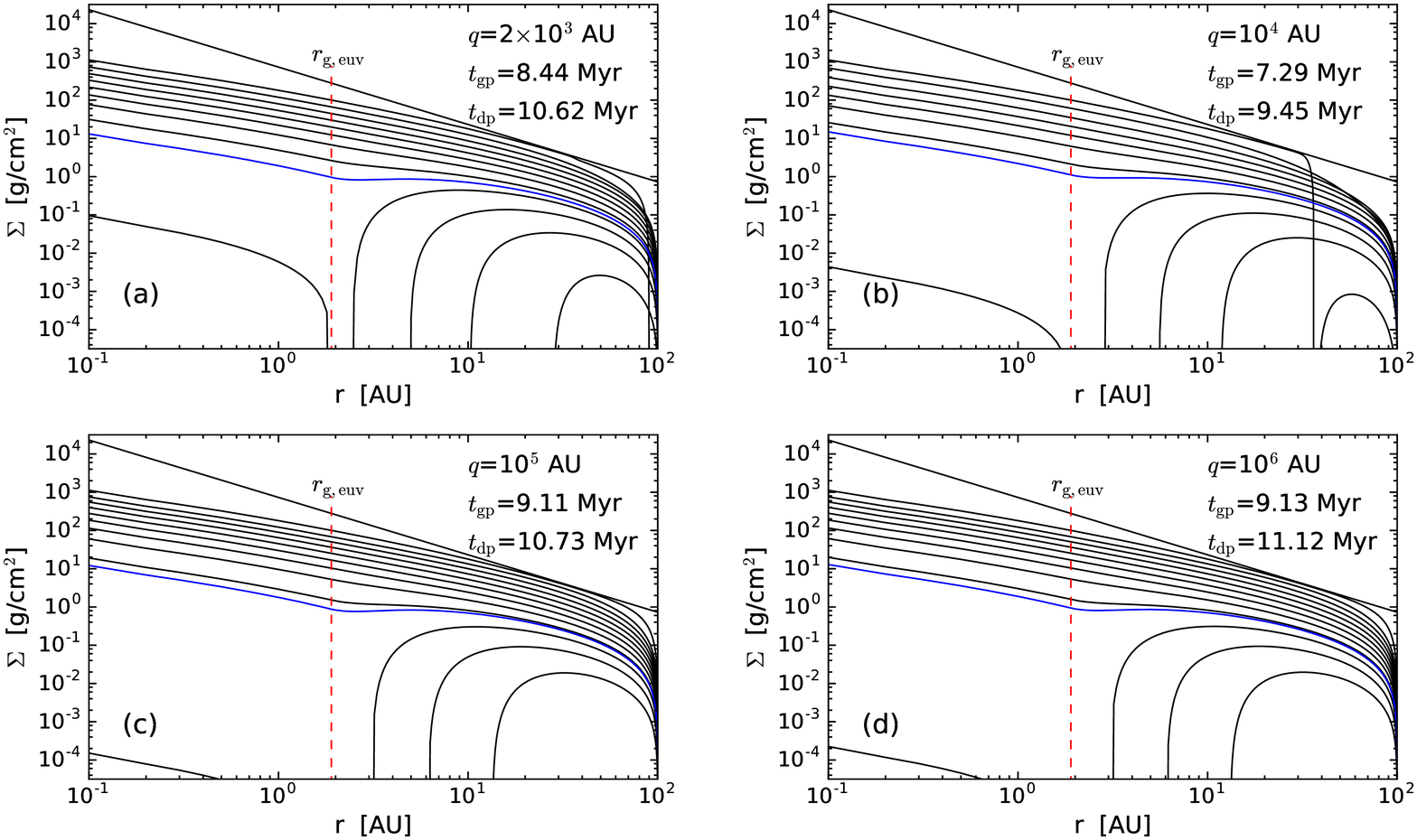}
	\caption{Similar with Figure \ref{Figure 9}, the snapshots of the surface density are plotted every 1 Myr, from top to bottom. The peri-center distance $q$ is 2$\times 10^{3}$ AU, $10^{4}$ AU, $10^{5}$ AU and $10^{6}$ AU. While $e$=10, $inc=0.5\pi$, and $T_{\rm  eff}=19000$ K. The duration of the flyby with $q=10^{6}$ AU is set as 10 Myr \label{Figure 10}} 
\end{figure*}

After setting $inc=0.5\pi$ and $e=10$, we change $q$ to investigate the influence on disc evolution due to different peri-center distance. In figure \ref{Figure 10}, four panels from top to bottom represent simulation results with prei-center distance $q=2000$, $10^{4}$ and $10^{5}$ and $10^{6}$ AU, respectively. When $q=10^{4}$ AU, disc evolution is influenced most obviously. As mentioned in section 3.2, according to Equation \ref{23} and Figure \ref{Figure 4}, the mass loss originated by external star's photo-evaporation will reach the maximum when $q\sim 10^{4}$ AU. If $q$ become smaller, the star will go fast and spend fewer time to go through FUV dominated region, and the influence on disc evolution becomes smaller. If $q$ becomes larger than $10^{4}$ AU, a smaller fraction of orbit is inside FUV region, or the star can not across FUV dominated region when $q>d_{\rm  max}$. Therefore the influence of external star becomes smaller. In the case of $q=10^{4} $ AU, the gap-open will becomes 7.29 Myr, while the disc disperse after $\sim$ 10 Myr. When $q=10^{5} $ AU, the disc evolution is nearly the same with that in Figure \ref{Figure 5}, which ignore the external star. Therefore if $q>10^{5}$ AU the photo-evaporation of external star can be ignored for stars with $T_{\rm  eff}$. In the case of $q=10^{6}$ AU, we set the flyby duration as 10 Myr(see in Appendix B), while the assumption of a 2 Myr flyby is also suitable as the external EUV mass loss due to a 10 Myr and 2 Myr are all smaller than one percent of initial disc mass.

To summarize the correlation between the disc evolution and the orbital parameters in this section, we can conclude that:
\begin{enumerate}
	\item 
	For typical external O, B stars with temperature $\sim 19000 $ K, photo-evaporations effects mainly depends on peri-center, rather than the eccentricity or inclination. Although smaller eccentricity and larger inclinations lead to larger mass loss and result in gas depletion between 9-10 Myrs.
	
	\item
	Flyby stars in the orbit with inclination $inc \ge 0.25 \pi$, lead to few changes with disc evolution. While in the case of coplanar flybys, i.e. the flyby orbit with very low inclination $\sim 0$, the gap-opening time $t_{\rm  gp}$ and disc dispersal time $t_{\rm  dp}$ will be put afterward $\sim 1 $ Myr, compared with the cases of $inc=0.25\pi$ and $0.5\pi$.
	
	\item 
	Flyby stars in the orbit with peri-center $q \sim 10^4 $ AU, will influence the disc evolution most obviously. Note that closer flybys of the same flyby stars can not accelerate the disc dispersal or gap-opening time. 
	
	\item
	For single flyby events with a typical B star with $T_{\rm  eff} \le 19000 $ K, in typical range of orbital parameters, i.e. $e \in [3,100]$, $inc\in[0,0.5\pi]$, $q\in[2 \times 10^{3}, 10^{5}]$ AU, the disc evolution can be changed. The gap-opening time is after 7 Myr, while disc dispersal time is larger than $\sim 9 $ Myr. Such a disc evolution timescale is typical for proto-planetary disc. Such the single flyby events can hardly influence the planet formation and orbital architecture around host stars due to photo-evaporation. 
	
\end{enumerate}

\subsection{The beginning time of flybys}
\begin{figure*} 
	\centering
	\includegraphics[width=0.9\linewidth]{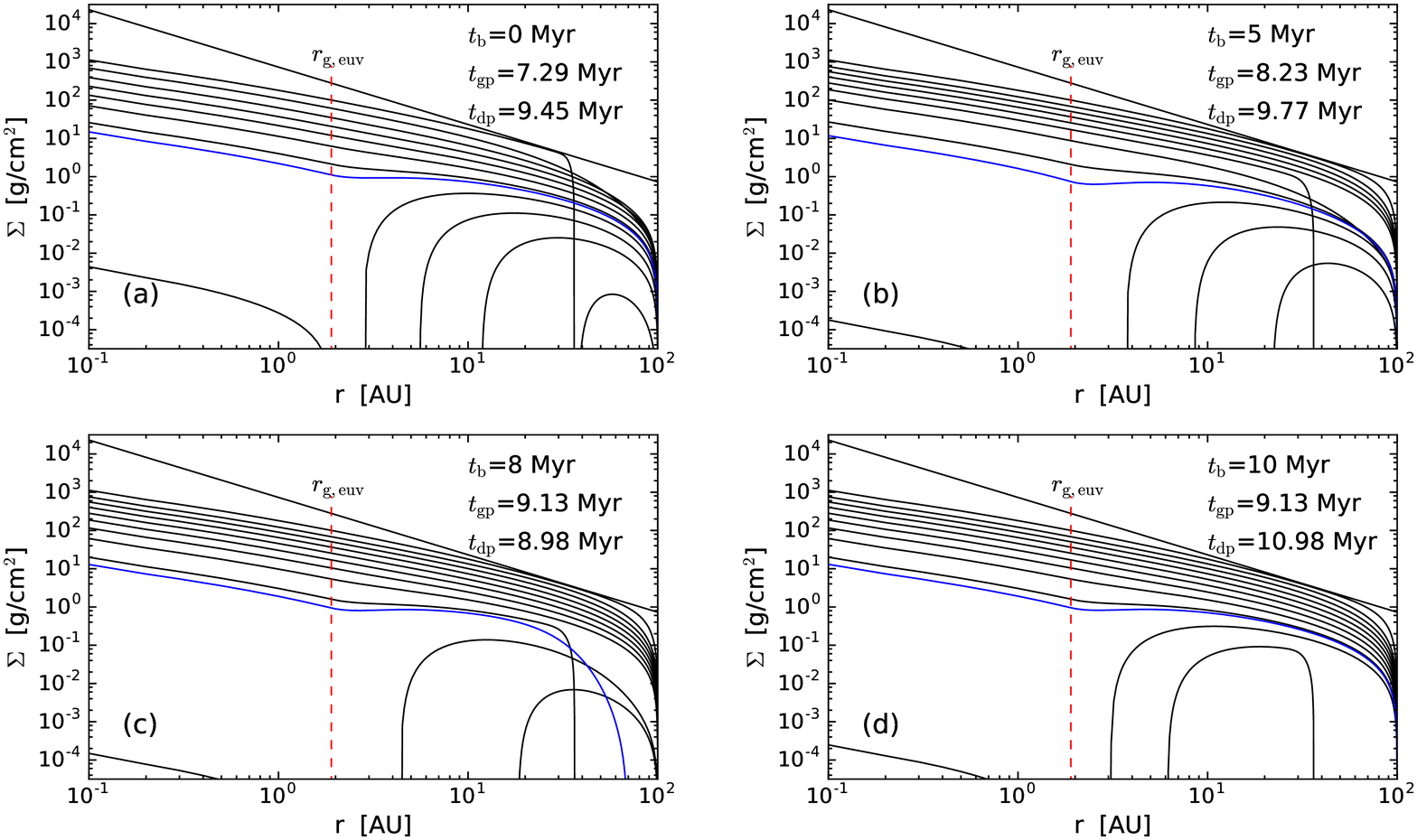}
	\caption{The disc surface density evolution with external stars with $T_{\rm  eff}=19000$ K. The flyby orbit is fixed as $inc=0.5\pi$, $e=10$ and $q=10000$ AU. The time when the flyby begins is set as follows: $t_{b}$=0 Myr, 5 Myr, 8 Myr, 10 Myr. The red lines represent $r_{\rm  g,euv}$, the gravitational radius where gas temperature is heated to $10^{4}$ K by EUV photo-evaporation. The blue lines represents the surface density profile when the gap is beginning to open. \label{Figure B1}} 
\end{figure*}
\begin{figure*} 
	\centering
	\includegraphics[width=0.9\linewidth]{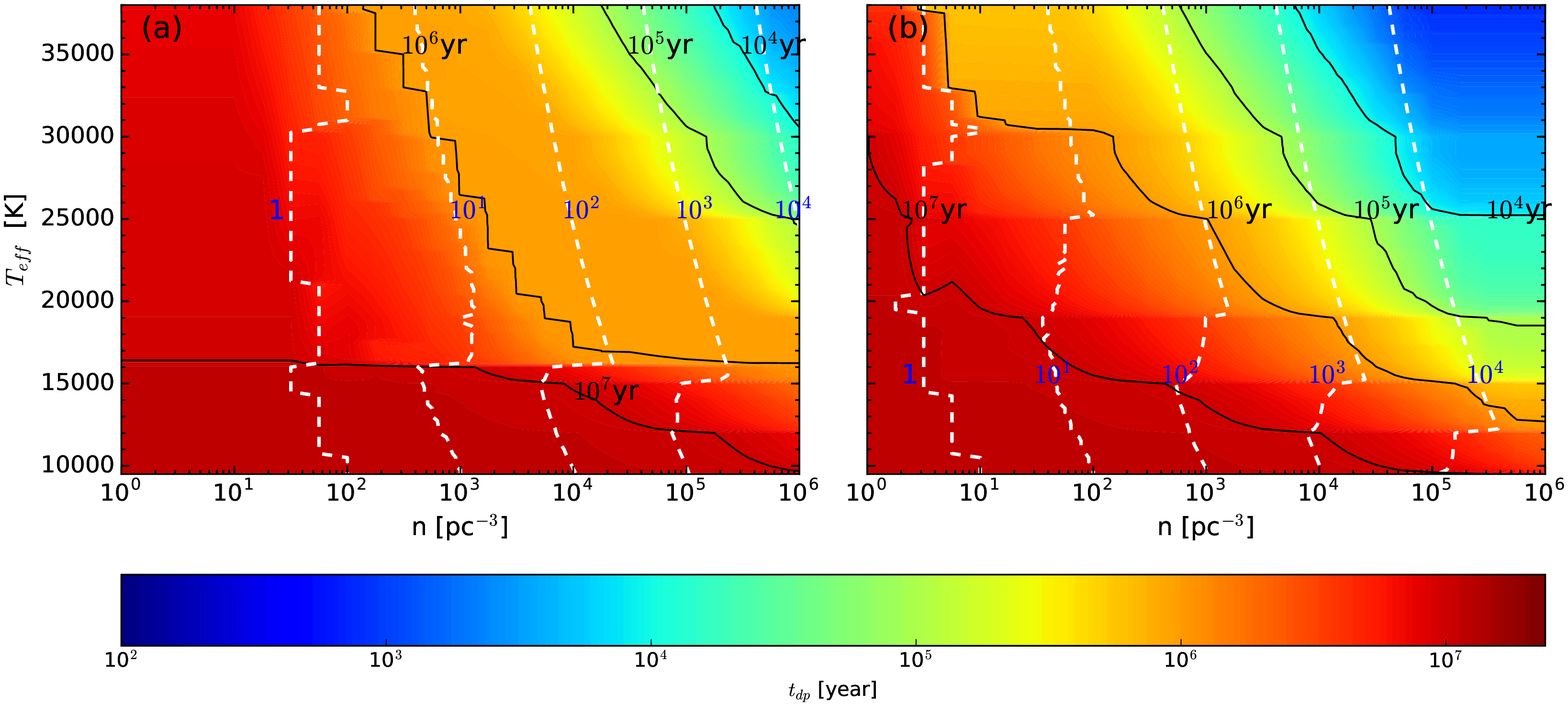}
	\caption{The the dispersal timescale of disc varies with effective temperature $T_{\rm  eff}$ of flyby stars and stellar density $n$. X axis is the star density in the surrounding environments. Y axis represents the effective surface temperature of the flyby stars. The color bar represents the dispersal timescale of the disc. Panel(a) and (b) show two different cases with peri-center distance $q=10^{4}$ AU and $q=10^{5}$ AU, respectively. Other parameters are the same for both panels. Black solid lines represent the different dispersal times varies in $10^{4}$ yr, $10^{5}$ yr, $10^{6}$ yr and $10^{7}$ yr. While white dashed lines represent the number of flyby events during disc dispersal timescale, i.e. $10^{0}$, $10^{1}$, $10^{2}$, $10^{3}$, $10^{4}$, from left to right, respectively. \label{Figure 11}}
\end{figure*}
In all the simulations above, we assume that the flybys begin at the initial time, i.e. t=0 Myr, and the duration of the flybys are 2 Myr. To investigate the influence of flyby events at different disc evolution stage, we set the different  beginning time of flyby event $t_{\rm b}$.
In this subsection, we fix the inclination as 0.5$\pi$, eccentricity $e=10$ and pericenter distance $q=10000$ AU. The effective temperature of external star is set as $T_{\rm eff}=19000$ K. The time when the flyby begins is set as follows: $t_{\rm b}$=0 Myr, 5 Myr, 8 Myr, 10 Myr. The other parameters are all the same as the case in Figure \ref{Figure 5}.

In figure \ref{Figure B1}, when the external star reaches the pericenter i.e. one million year after the beginning of the flyby, the disc outer edge shrinks to the gravitational radius $\sim \beta r_{\rm g,fuv}\sim 36$ AU. Then the viscous evolution of the disc recovers the disc outer edge that was truncated by the external star.

If we consider the influence on the gap open time $t_{\rm gp}$, we can see that the later the flyby event takes place, the later the photo-evaporation gap opens. Considering that the influence of the external FUV photo-evaporation is rather efficient, so when the external star approaches to the pericenter, the outer disc edge always shrinks to $\sim 36 $ AU. I.e. the gas material will be swept out where $r>36$ AU. Basically, the surface density is declining due to the viscosity evolution and internal photo-evaporation. So naturally, the earlier the flyby takes place, the more gas in outer disc can be swept out, thus results in smaller disc mass, and thus smaller mass loss due to viscosity. Therefore, earlier opening gap is achieved (see section 4.2).

When it comes to the disc dispersal time $t_{\rm dp}$, it seems that a little difference compared with $t_{\rm gp}$. As we can see in panel (c), if $t_{\rm b}$=8 Myr, it's surprise that $t_{\rm dp}$ is the shortest in all four cases.  We have defined the disc dispersal time as when the outer disc mass ($>36$ AU) is lower than one percent of the initial outer disc mass (approximately $4\times 10^{-5}$ M$_\odot$). Specifically, considering the recovery of the outer disc edge after flyby events, if the disc mass outside 36 AU can quickly accumulate to over $4\times 10^{-5}$ M$_\odot$, the outer disc mass decreases due to viscosity. As soon as the outer disc mass becomes less than $4\times 10^{-5}$ M$_\odot$, the disc dispersal time. However, In case of panel (c), when the flyby occurs in the late stage of disc, the outer disc can hardly achieve a mass of $4\times 10^{-5}$ M$_\odot$ due to viscosity after flyby, because of the low mass of total disc. Thus the disc dispersal time is adopted when the outer disc is clear up by the external photo-evaporation, i.e. the external star arrive near pericenter at $\sim t_{\rm b}+1$ Myr. Similarly, in panel (d), $t_{b}$=10 Myr, the flyby begins after the photo-evaporation gap opens. The disc mass is very low and disperse when the external star truncates the outer disc edge at 36 AU.

In summary, flyby events in early stage of disc can influence the disc evolution, e.g. when $t_{\rm b}\le 5$ Myr, the later the flyby begins, the later gap opens and the later disc disperses. Flyby events in the late stage hardly influence the disc evolution, e.g. $t_{\rm b}\in [8,10]$ Myr, then $t_{\rm gp}$ is almost the same as case with no flyby such as Figure \ref{Figure 5}, and $t_{\rm dp} \approx t_{\rm b}+1$ Myr. I.e the external star arrives at peri-center and the mass loss due to external FUV photo-evaporation becomes much more efficient than viscosity, the disc mass outside 36 AU decreases quickly and finally disperses around $t_{\rm b}+1$ Myr.

\section{Proto-planetary disc lifetime in clusters: multi-flyby effects}
As we conclude in section 3 and 4, a single flyby event can not influence disc lifetime significantly in most cases, even if the external star is very hot. Considering the fact that stars form in clusters or groups, the young stars with proto-planetary disc are usually in crowded environments. In such environments, close encounters with other stars are frequent. Therefore one star will encounter multiple flyby events. The time-scale for a given star to experience an encounter with another star, with a peri-center distance $q$, is given by \citet{2007MNRaS.378.1207M}, i.e. 
\begin{equation} \label{26}
\tau_{\rm  enc}\simeq 3.3 \times 10^{7} \rm{yr} 
\left(\frac{100 \rm pc ^{-3}}{\emph{n}}\right)
\left(\frac{\nu}{\rm １km s^{-1}}\right)
\left(\frac{10^{3} \rm AU}{\emph{q}}\right)
\left(\frac{\rm M_{\rm  \odot}}{\emph{m}_{\rm  t}}\right),
\end{equation}
where \emph{n} is the star number density in the cluster, \emph{v} is the mean relative speed at infinity of the objects in the cluster, and $m_{\rm  t}$ is the total mass of the two flyby stars. Assuming that a less dense cluster with $n$ = 100 pc$^{-3}$, $v$ = 1 km s$^{-1}$, $q$ = 10000 AU, and $m_{\rm  t}$ = 10 M$_{\rm  \odot}$, we can get the time-scale of close encounters in such cluster is 0.33 million years. In other words, the frequency of the close encounter take place is 3 per million years. As the lifetime of a typical proto-planetary disc is around 1--10 million years, the disc usually experience many flyby events.

Due to mass segmentation effect, in the dense core of the clusters there are usually some massive stars. The radiation of these stars, including EUV and FUV, is dramatically strong \citep{2007MNRAS.376.1350C}. E.g. the star Orion $\theta^{1}$ in Trapezium open cluster. If a star approaches to these massive star several times, the strong EUV and FUV radiation of the massive stars will influence the proto-planetary disc more obviously than the case of single flyby events. In this section, we will develop a simple multiple flyby model and achieve the disc dispersal time in different environments.

\subsection{Multiple flyby models and simulation results}
In order to get the correlation between radiation of external stars, star density and the dispersal timescale of gaseous disc, we use a simple multiple flyby model. We set the parameters of central star as follows: the initial disc mass $ M_{\rm  d0}$=0.01 $\rm M_{\rm  \odot}$, central star mass $ M_{\rm  c}$=0.5 $\rm M_{\rm  \odot}$. During a typical disc life time of 10 Myr, the flyby events will occur $N_{\rm  flyby}$ times, where

\begin{equation} \label{27}
N_{\rm  flyby}=[\frac{10 \rm Myr}{\tau_{\rm  enc}}]+1.
\end{equation}
$\tau_{\rm  enc}$ can be estimated via Equation \ref{26}. The square brackets is the integer-valued function, to return the integer part of a real number. 
To obtain a confidential results, it's better to do N-body simulation to give the distribution of orbital parameters and external stellar mass during multiple flybys. Since the distribution is sensitive to the initial conditions of clusters, here we only take the same orbital parameters and the same external star for all flyby events. We only consider two cases with different pericenter distance $q=10^{4}$ AU or $10^{5}$ AU. Such a large $q$ will have tiny gravitational effect on the disc \citep{2006apJ...642.1140O}. We assume that  the intruders will not be captured by the center star to form binary systems. According to the Equation \ref{26}, both  mass of externals star and the star density in clusters will influence the flyby timescale. Therefore, we choose the temperatures of the most massive stars (from 9500 to 38000 K) in the clusters as one free parameter, which decide the masses and the FUV and EUV flux according to Table \ref{Table 1}. Another free parameter is the density of star $n$, which represent the different dense of clusters, e.g. open or globular clusters. The inclinations are fixed as $inc=0.5\pi$, while the eccentricities are set as $e=10$. According to the vis-viva equation,

\begin{equation} \label{28}
q = 4\pi^2 \frac{\mu}{v^{2}}\left( e-1 \right)
\end{equation}

We can estimate the initial relative velocity of the external star 0.4 $\sim$ 6 km s$^{-1}$, according to the parameters we set above. The value is consistent with the stellar velocity dispersion in open clusters($\sim 1 $ km s $^{-1}$)\citep{1989AJ.....98..227G} and the stellar velocity dispersion in globular clusters(1 $\sim$ 25 km s$^{-1}$)\citep{1993ASPC...50..357P}. 

During each flyby event, we calculate the photo-evaporation of external stars, only if it's larger than the photo-evaporation of central star. The time of peri-center of different flyby events distribute in 10 Myr uniformly.

In figure \ref{Figure 11}, the black solid contour lines show the disc dispersal time with different $T_{\rm  eff}$ and the white dashed contour lines show the different number of flyby events before the dispersion of gas disc. In panel (a), $q=10^{4}$ AU is adopted, while in panel (b) $q=10^{5}$ AU is adopted. These two panels have similar structures. Both panels can be divided into three regions according to star density $n$, i.e. filed star with $1<n<10^{2}$ pc$^{-3}$, open clusters with $10^{2}<n<10^{5}$ pc$^{-3}$, and globular clusters $10^{5}<n<10^{6}$ pc$^{-3}$. As star density increases and the flyby events become frequent, which leads to stronger photo-evaporation of gaseous disc, therefore the disc disperse in early time.

In panel (a), when the effective temperature of external star $T_{\rm  eff} \lesssim 16000$ K, the FUV dominated region is always inside $10^{4} $ AU(see Table \ref{Table 1}). I.e. the external star with $q=10^{4}$ AU only influences the gaseous disc via EUV flux. Thus there is a horizontal structure of dispersal time $t_{\rm  dp}=10$ Myr around $T_{\rm  eff}=16000 $ K. The area above 16000 K means the FUV photo-evaporation begin to work. Both EUV and FUV dominate the mass loss of disc and determine the $t_{\rm  dp}$. While in the area under 16000 K, only EUV photo-evaporation and viscosity works. In the bottom region, where the dispersal time $t_{\rm  dp} \geq 10^{7}$ yr is similar to the case without external star as shown in figure \ref{Figure 5}. As mentioned in section 4, the mass loss due to EUV photo-evaporation is small. In this area, The external star's photo-evaporation is very limited and can be neglected. The disc dispersal timescale is determined by disc viscosity, therefore $t_{\rm  dp}$ varies little in the viscosity dominated region. 

In the area with $T_{\rm  eff}<16000 $ K and $t_{\rm  dp}<10^7$ yr, because the stellar density becomes larger and larger, the EUV photo-evaporation becomes more and more obvious due to multiple flybys, and begins to influence the disc dispersal timescale. In this EUV dominated area, with the increase of $n\in [10^{3}, 10^{6}] $ pc$^{-3}$, external star with smaller $T_{\rm  eff} \lesssim 16000$ K can affect the disc dispersal. I.e. the $t_{\rm  dp}$ will be put forward in dense open clusters or globular clusters, because of much frequent flyby events. Note that $t_{\rm  dp} > 1 $ Myr, even in the extreme case of $n=10^{6}$ pc$^{-3}$ and $T_{\rm  eff}=15000 $ K. It is possible to form gas giant in the first 1 Myr, only if the massive core can be formed very quickly. However, the gas giant is probably not as massive as Jupiter, if the gas deplete fast. 

In the area where $T_{\rm  eff} > 17000 $ K, the dispersal timescale $t_{\rm  dp} \lesssim 10 $ Myr because of large mass loss due to FUV photo-evaporation. However, stars in clusters with $n<100$ pc$^{-3}$ can sustain a disc longer than 1 Myr, even the flyby stars are extremely hot. Therefore, stars in clusters with low densities ($<$ 100 star per cubic pc) can hardly influence the formation of gas giants if a solid core can be formed quickly. In the top right corner of the diagram, where $T_{\rm  eff} \in [25000, 38000]$ K and $n\in [10^{4},10^{6}]$ pc$^{-3}$, the disc dispersal time $t_{\rm  dp} \lesssim 10^{5} $ yr. It means in the environments of very dense clusters with extremely hot stars, the gaseous disc can only survive for a very short timescale ($< 0.1$ Myr) and gas giant hardly form because of the gas deficiency.

Panel (b) shows similar figures with panel (a). When the effective temperature of external star $T_{\rm  eff} \lesssim 30000$ K, the FUV dominated region is always inside $ \sim 10^{5} $ AU(see Table \ref{Table 1}), I.e. the external star with $q=10^{5}$ AU only influences the gaseous disc via EUV flux. The area above 30000 K means the FUV and EUV photo-evaporation dominates the mass loss of disc and determines the $t_{\rm  dp}$, while in the area under 30000 K, only EUV photo-evaporation and viscosity works. 

Similar with panel (a), the area with $t_{\rm  dp} \geq 10^{7} \rm $ yr represents that the viscosity dominates the disc dispersal. For stars with higher temperature, there is a smaller viscosity dominated area because of the enhancement of EUV photo-evaporation. The other area with $T_{\rm  eff}<30000$ K is dominated by EUV photo-evaporation. Compared with the similar area in panel (a), the EUV dominated area in panel (b) is enlarged a lot. It is because in the case of $q=10^5 $ AU in panel (b), more hot stars with $T_{\rm  eff}=30000 $ K are included, which have a much stronger EUV than cooler stars. With the same stellar density, hotter stars lead to larger disc mass loss, and $t_{\rm  dp}$ is reduced more obviously than that in panel (a). Another reason is, the flyby events with larger $q$ occur more frequently than those with smaller $q$.

In the both FUV and EUV dominated area $T_{\rm  eff} > 30000 $ K, with the increase of stellar density $n$, the frequency of the flyby events increases. Therefore the disc mass loss caused by photo-evaporation increases and the dispersal time $t_{\rm  dp}$ decreases consequently. It's similar with panel (a). The difference between panel (a) and (b) is that, the dispersal timescale $t_{\rm  dp}$ in panel (b) is shorter that in panel (a). It's the obvious results because of the larger $q=10^{5} $ AU in panel (b), i.e. the flyby events take place more frequent than those in panel (a) according to Equation \ref{26}. More specifically, in panel (a), the gas disc disperses in 1 Myr, after 10 flyby events, if the external stars have the temperature of 20000--25000 K, or after several flyby events if the external stars are very hot ($> 30000$ K). In panel (b), disc disperse in 1 Myr, after tens of flybys with hot stars (20000--25000 K), or after several flybys with a very hot star ($>$ 30000 K).

\subsection{The disc lifetime in clusters: observations}
In our paper, we consider the EUV and FUV photo-evaporation from the external stars in the flyby, and then develop the model into cases in clusters. In section 5.1, we have shown that in young open clusters(1<n<100 [pc$^{-3}$]), if there is not a massive central star whose effective temperature is larger than 16000K, the disc dispersal time scale is close to several million years. While massive central stars may rapidly photo-evaporated the disc surround it, depending on the temperature of stars in the fly-by and the peri-center of flyby events. 

According to \citep{2001apJ...553L.153H} one-half of the stars within the clusters lose their discs in $\le$ 3 Myr. \citep{2008apJ...686.1195H} has indicated that the low frequency of primordial discs reflects the fast disc dissipation observed at 5 Myr in $\gamma$ Velorum stellar clusters. Based on the survey in young $\sigma$ orionis cluster, \citep{2007apJ...662.1067H} suggests that external photo-evaporation is not very important in many clusters because of the lack of massive central stars. All these observation is consistent with our results, i.e. the most disc dispersal timescale is between 1 and 10 Myr, in open clusters with $n<10^3$ pc$^{-3}$. The most recent exoplanet survey in M67 open clusters, shows an occurrence of gas giant $\sim$ eighteen percent, which is similar with the occurrence rate of gas giants around field star \citep{2017a&a...603a..85B}. In such a cluster like M67, with $n < 100$ pc$^{3}$ \citep{2005MNRaS.363..293H}, the disc can survive several Myr, like the disc around field stars. Therefore, the formation of gas giants is probably similar to the field stars.

There are also some other photo-evaporation work to explain the observations in open clusters. \citet{2006apJ...641..504a} shows that with the median external FUV flux $G_{\rm  0} \sim 900$(about 1.4 ergs cm$^{-2}$ s$^{-1}$), in typical size of open clusters, the external FUV will not influence the disc evolution much. Although their model do not consider the cases of massive stars in clusters. Our model consider in more details. E.g. the orbit parameters during a single flyby. Besides young open clusters, we also extend our model from field star to the globular clusters, and get the result that disc is hard to survive in such extreme radiation surrounding in globular clusters. 
\section{Conclusions}
Since planets formed in proto-planetary gaseous disc, the evolution of gaseous disc is crucial for planet formation, especially for gas giants like Jupiter. The lifetime of the gaseous disc can determine whether the gas giant can form or not, and how large the gas giant can grow due to core accretion scenario. The Photo-evaporation is an important mechanism during disc evolution and lead to different mass loss rate of the proto-planetary disc. 

In this paper, we use a general one dimension disc model (section 2), including EUV and FUV photo-evaporation, to investigate the disc evolution due to different flyby stars in different hyperbolic orbits. In section 3, we analysis the mass loss during single flyby event due to external star in two cases, which depend on the orbital parameters as shown in Equation \ref{20} and \ref{22}. Using one-dimension diffusing equation, we simulate the disc evolution due to both viscosity and photo-evaporation of both external and central stars for single flyby events in section 4. The timescales for gap-opening and depletion of gaseous disc due to different flyby events are obtained. To model the real star birth environments, we simulate disc evolution via a simple multi-flyby model in section 5.


We list the main conclusions in this paper as follows:
\begin{itemize}
	\item Using atlas9 model, we calculate the EUV and FUV flux of different stars with effective temperature from 9500 to 38000 K. The FUV photo-evaporation regions for different external stars are estimated as shown in Table 1. 
	
	\item We derive analytical equations of total mass loss due to external star with different parameters, including external ionization flux, eccentricity $e$, inclination $inc$ and peri-center distance $q$ of hyperbolic orbit. In the case of EUV dominated photo-evaporation when $q > d_{\rm  max}$, single flyby lead to very few mass loss (Equation \ref{19}). Considering the FUV photo-evaporation region, the FUV photo-evaporation is much more efficient than EUV, thus the mass loss of disc during flyby mainly depends on how long the star go across the FUV dominated region.  In the case of $T_{\rm  eff}=19000 $ K, external star with peri-center $\sim 10000 $ AU lead to the maximum mass loss as shown in Figure \ref{Figure 4}.
	
	\item Using the alpha-disc model with $\alpha=0.001$, and consider the photo-evaporation of central star, we find that the ionization flux of external star and the peri-center distance $q$ are crucial for the disc evolution after flybys, while the inclination and eccentricity influence the disc evolution limitedly. Only Stars with $T_{\rm  eff} > 15000 $ K in an orbit with peri-center $q=10000$ AU can reduce the disc lifetime. For the single flyby cases of $T_{\rm  eff}\in [9500,38000] $ K in Figure \ref{Figure 6}, or $q\in[2000,10^6]$ AU in Figure \ref{Figure 10}, the lifetime of gaseous disc is $>7$ Myr, which is typical for proto-planetary disc, and can not significantly restrain the formation of gas giants. 
	
	\item By changing the dimensionless viscosity alpha and ionization luminosity $\Phi_{\rm c}$(in internal photo-evaporation and viscosity evolution case), we find that, with a larger $\alpha=0.01$ the disc can disperse at less than 2 Myr, which is well consistent with the observations that protoplanetary gaseous disc can disperse in 3 Myr. While increasing the central ionization luminosity to $10^{42}$ s$^{-1}$, the gap opens at an earlier time. We also find that the disc viscosity, instead of the internal ionization luminosity, dominates the outer disc evolution after gap opens in our model.
	
	\item During disc evolution, the gap opens when the mass transfer at $r_{\rm  g,euv}$ due to viscosity is less than the mass loss due to photo-evaporation. We derived an criterion to estimate the gap-opening timescale $t_{\rm  gp}$, which is consistent with simulations(Figure \ref{Figure 7}). According to our simulations, the gap usually opens when the disc is nearly depleted as shown in Table \ref{Table 2}. Since the gaseous disc is less massive, the gap-opening can hardly influence the migration of planets in the disc.
	
	\item Flyby events in early stage of disc can influence the disc evolution, e.g. when $t_{\rm b}\le 5$ Myr, the later the flyby begins, the later gap open and the later disc disperses. while flyby events in late stage of disc can influence the disc evolution limitedly, e.g. $t_{\rm b}\in [8,10]$ Myr, $t_{\rm dp}$ is almost the same as the case with no flyby in Figure \ref{Figure 5}, and $t_{\rm dp} \approx t_{\rm b} +1$ Myr.
	
	\item Since flyby events occurs frequently in clusters. We investigate the photo-evaporation of gaseous disc during multi-flyby events. In both cases of $q=10^{4}$ and $10^{5}$ AU, there is an area with stellar density $< 100$ pc$^{-3}$ and no massive stars ($T_{\rm  eff}< 30000$ K) in the clusters, e.g. open cluster M67, where the gas dispersal timescale $t_{\rm  dp}$ is between 1--10 Myr. I.e. the disc lifetime is typical for the formation of gas giants. For globular clusters with stellar density $>10^5 $ pc$^{-3}$, where there usual are some massive stars, the disc will depleted in 0.1 Myr, therefore gas giants can hardly formed in such an environments.
	
	
	\item When the external star($T_{\rm  eff}>16000$ K see figure \ref{Figure 6} and \ref{Figure 11}) is coming close to the peri-center, the outer disc edge will shrink, which is similar to the surface density profile shown by \citep{2003ApJ...582..893M}. When the external star is far away from the center star, the external photo-evaporation can be negligible, the disc evolution is dominated by the viscosity and central photo-evaporation which is similar to the UV-switch model \cite{2001MNRAS.328..485C,2006MNRAS.369..216A,2006MNRAS.369..229A}, when the $\dot M = 2*\dot M{\rm c,euv}$, the disc will open a gap at the soften gravitational radius $\beta r_{\rm  g,euv}$, and the inner disc will dissipate at the viscous times scale $ \sim 0.1$ Myr based on our parameter setting. 
\end{itemize}

Previous external photo-evaporation models, e.g. \citep{2003ApJ...582..893M,2013apJ...774....9a,2018ApJ...853...22X} do not consider the change of distance between external star and central star, where our model mainly focuses on. They conclude that the disc dispersal timescale should become much less than 10 Myr. Considering a smaller gaseous disc, \citep{2013apJ...774....9a} obtain the gas depletion timescale of 1-3 Myr, when only FUV radiation with $G_0=300-30000$ works, which is $10^2-10^4$ higher than the interstellar value. Our results show a flyby events with external star with $T_{\rm  eff}=38000 $ K(with $G_0 \sim 30000$ at $d=10^5$ AU), including both EUV or FUV depends on the distance. The depletion timescale of gaseous disc is put forward $>$ 7 Myr, but larger than the previous works.

In this paper, we only consider the flyby with peri-center larger than 2000 AU, to avoid the influence of gravitational effects. Recently, \citet{2014MNRAS.441.2094R} considers both the hydrodynamical evolution of the discs around their natal stars, as well as the dynamics of the stars in clustered environments. As discs evolve due to viscosity, encounters become more and more important, it will truncate the discs and deprive of the outer potions because of the dynamical tidal effect. For close flybys with $q\sim 10^{2}$ AU, the dynamical effect of external star is significant. While for wide flyby events with $q\ge 10^{4}$ AU, the mass loss due to dynamical effect of flyby can be neglected(see figure 3 in their paper). Their results are consistent with our assumption that, when $q\ge 10^{4}$ AU, the gravitational effects can be ignored. They predict that disc sizes are limited by encounters at stellar densities exceeding $\sim 2$ - $3 \times 10^{3}$ pc$^{-2}$. In our work, we do not consider the gravitational effects, but mainly focus on the external photo-evaporation in clustered environments. And our results show that, in globular clusters with high stellar density $>10^{5}$ pc$^{-3}$, disc lifetimes will be probably reduced to less than 1 Myr. 

In our model, we do not consider the binary cases, although the binary stars are common in universe \citep{2010apJS..190....1R}. Recently, \citet{2018MNRaS.473.5630R} study the evolution of proto-planetary discs in binaries. They focus on the X-ray photo-evaporation of each star in binary and also the tidal effect between companion star and disc. They conclude that the disc around close binary star disperse quickly than single stars due to tidal effect between the companion and the disc. Based on the results of disc in binary, we can deduce that the close flyby with small peri-center will clear the disc more quickly. However, these flybys are rare according to Equation \ref{26}, thus only a few disc may be influenced by the close flyby events accidentally.

Note that the massive star is not the only source of EUV and FUV radiation in clusters. Both the accretion disc around Black hole, and accretion disc in compact binary can produce X-ray emission. These additional radiation can strengthen the photo-evaporation and then accelerate the evolution of a proto-planetary disc. Contrarily, in a young embedded cluster with plenty of gas, the strong extinction of EUV and FUV can reduce the mass loss due to photo-evaporation and protect the proto-planetary disc. 

We adopt a very simple multiple flyby model to investigate photo-evaporation in clusters. As mention before, to get a more reasonable parameters of flyby, it's better to do N-body simulations to obtain the distribution of the mass, peri-center, eccentricities, and inclinations of flyby stars in clusters. However, these distributions are sensitive to the cluster model, especially the initial conditions and IMF of clusters. The results could be more specific, if we can get the distribution of flyby parameters in some clusters.  

We can explain the gas giant formation is hard in globular clusters because of short-lived gaseous disc, which is consistent with the null survey results in globular clusters \citep{2017aJ....153..187M}. Our results in this paper predict that the stars in open clusters without very massive stars can sustain a disc similar to the field star, thus the planet occurrence is probably the same as that of field star. It's not constrained well by observation because of few planets samples in open clusters. Further surveys in different clusters with or without massive stars can provide clues to investigate the influence of planet formation rate in clusters. 


\section*{Acknowledgements}

This work is supported by the National Natural Science Foundation of China (Grant No.
11503009), and the Fundamental Research Funds for the Central Universities(14380016,14380018). NSFC(Grant No. 11333002, 11673011) and the Key Development Program of Basic Research of China (No. 2013CB834900) also support this work.



\bibliographystyle{mnras}
\bibliography{reference}




\appendix 

\section{Disc mass loss originated by external stellar gravity}
According to \citep{2006apJ...642.1140O}, the relative disc mass loss due to gravitation during parabolic, coplanar encounters can be approximated by
\begin{equation} \label{Equation A1}
\frac{\Delta M_{\rm  d}}{M_{\rm  d}}
=\left( \frac{m_{\rm  ex}}{m_{\rm  ex}+0.5m_{\rm  c}} \right)^{1.2} 
\ln \left[ 2.8\left( \frac{q}{r_{\rm  d}} \right)^{0.1} \right]
\times \exp 
\left\{
-\sqrt{\frac{m_{\rm  c}}{m_{\rm  ex}+0.5m_{\rm  c}}}
\left[
\left(\frac{q}{r_{\rm  d}}\right)^{1.5}-0.5
\right]
\right\}
\end{equation}
Where $M_{\rm  d}$ is the disc initial mass. $m_{\rm  ex}$ is the mass of the external star. $m_{\rm  c}$ is the mass of central star. $q$ is peri-center distance. $r_{\rm  d}$ is the disc edge. This equation is the expanded expression according to \citep{2005a&a...437..967P}. It's limited to close encounters with star mass ratios in the interval 0.1<$m_{\rm  ex}/m_{\rm  c}$<625. If we give the parameter setting such as $m_{\rm  ex}\sim 40 $M$_{\rm  \odot}$, $m_{\rm  c}=0.5$M$_{\rm  \odot}$, $q=2000$ AU, $r_{\rm  d}=100$ AU. We can estimate the mass loss ratio $\frac{\Delta M_{\rm  d}}{M_{\rm  d}}\sim 10^{-5}$. For an hyperbolic orbit with eccentricity $e=10$, the mass and angular momentum transfer is only 20 per cent of that in a parabolic encounter\citep{2005a&a...437..967P}. Apart from that, in our paper, the initial disc surface density distribution is proportion to $r^{-1.5}$ which is steeper than that in \citep{2006apJ...642.1140O} with $r^{-1}$. With denser inner disc, and for the sake of the gravitational capture of central star,  fewer mass and angular momentum transfers in a close encounter. Thus, in this paper, mass loss due to the gravitational effect can be ignored, if we set the pericenter distance $q \ge 2000$ AU.

\section{Derivation of the analytical solution of external photo-evaporation}
The hyperbolic from of Kepler's equation is 
\begin{equation} \label{Equation B1}
N=e\sinh H-H
\end{equation}
Where H is the hyperbolic eccentric anomaly, and the quantity $N$ is similar to the mean anomaly of elliptic motion and is defined as 
\begin{equation} \label{Equation B2}
N=2\pi\sqrt{\frac{\mu}{(-a)^{3}}}(t-t_0)
\end{equation}
where $\mu=m_{t}/1$M$_{\rm \odot}$, $m_{t}$ is the defined as the total mass of the two stars. $a$ is the semi-solid axis of the hyperbolic orbit, which is negative(in unit of AU). $t_0$ is the initial time when the star is crossing the peri-center(in unit of yr). The gravitational constant $G=4\pi^{2}$ when we adopt the unit of M$_{\odot}$, AU and yr above. Since the symmetry of the orbit, we only need to integrate one half of the orbit. And for the convenience of calculation, we set $t_0$ is equal to zero. The distance between two stars can be written as:
\begin{equation} \label{Equation B3}
d=a(1-e\cosh(H))
\end{equation}
when the external star comes across the peri-center the relation becomes:
\begin{equation} \label{Equation B4}
q=a(1-e)
\end{equation}
assume that the angle between the vector of two stars and the norm vector of disc plane is $\theta$. Considering the transfer matrix from the orbit plane to the disc plane, we can obtain that
\begin{equation} \label{Equation B5}
\cos\theta = \left|\frac{e-\cosh H}{e\cosh H-1}\sin(inc)\right|
\end{equation}
Note that, the external star is at the peri-center initially and the $\cos\left(\theta\right)>0$, i.e $t_0=0$. It's obvious that when the external star comes across the disc plane, $\cos\left(\theta\right)=0$, i.e. there exits a $t_{\rm  c}$ satisfying the relation that $\cosh[H(t_{\rm  c})]=e$. 
Then we can substitute equation \ref{Equation B3} and \ref{Equation B5}  into the equation \ref{7} and \ref{17}, and use $f_{\rm  s}=|\cos \theta|$ instead of equation \ref{8} for simplicity, thus the $\Delta M_{\rm  euv}$ can be expressed as follows, 
\begin{equation} \label{Equation B6}
\Delta M_{\rm  euv} = 
\begin{cases}
2  D\left( \frac{t_{\rm  dur}}{2}\right) , & t_{\rm  dur}<2t_{\rm  c} \\
-2 D(\frac{t_{\rm  dur}}{2})  + 4 D\left( t_{\rm  c}\right), & t_{\rm  dur} \ge 2t_{\rm  c} 
\end{cases}
\end{equation}
where 
\begin{equation} \label{Equation B7}
D\left( H(t)\right) = C_{\rm  0}\sin(inc)\sqrt{\frac{q/1 \rm AU}{(e-1)\mu}}
\left[-\frac{H-2\sqrt{-1+e^{2}}\arctan\left(\frac{(1+e)\tanh(H/2)}{\sqrt{-1+e^{2}}}\right)}{e}\right].
\end{equation}
$t_{\rm  dur}$ is the duration of a flyby, and the factor 2 is considered for the symmetry of the hyperbolic orbit. The constant $C_{\rm  0}=2.3 \times 10^{-4}  \rm M_{\rm  \odot}(\frac{\Phi_{\rm i}}{10^{49}})^{1/2}(\frac{r_{\rm  d}}{100 \rm AU})^{3/2}$.

Considering the hyperbolic form of Kepler's equation 
\begin{equation} \label{Equation B8}
\sinh H=\frac{N+H}{e},
\end{equation}
we expand H as a series via Lagrange's expansion theorem, i.e.
\begin{equation} \label{Equation B9}
H =A+\frac{A}{B}+\frac{A}{2 B^{2}}\left(2-\frac{AN}{B}\right)+\cdots
\end{equation}
where $ A=arcsinh \frac{N}{e}=\log\frac{N+B}{e}$ and $B=\sqrt{N^{2}+e^{2}}$. Assuming that $e\gg 1$ and $t_0=0$, we can derive the expression of $H(t)$
\begin{equation} \label{Equation B10}
H=\log\left (kt+\frac{\sqrt{k^{2}t^{2}+\emph{e}^{2}}}{\emph{e}}\right)
\end{equation}
where $k=2\pi\mu^{0.5}(e-1)^{1.5} q^{-1.5}$. Assuming $e \gg 1$, $k\approx2\pi\mu^{0.5}e^{1.5}q^{-1.5}$, and $H$ becomes
\begin{equation} \label{Equation B11}
H\approx\log(2\pi\mu^{0.5}e^{1.5}q^{-1.5}t)
\end{equation}
With the typical parameters with $\mu\ge 10 $, $e\ge10$, $q\ge10^{4}$ AU and $t\ge1$Myr, we can estimate $H\ge6.4$. According to Equation \ref{Equation B10}. The term in $\arctan$ in Equation \ref{Equation B7} is very close to 1. Thus Equation \ref{Equation B7} is simplified as
\begin{equation} \label{Equation B12}
\begin{aligned}
D\left(H(t)\right) \approx &C_{\rm  0}\sin(inc)\sqrt{\frac{q/1 \rm AU}{e \mu}}\left[-\frac{H(t)}{e} + 
\frac{\pi}{2}\right]
\end{aligned}
\end{equation}

Note that $D(H)$ is not convergent with the increase of $t$. It's reasonable that in Equation \ref{7}, the $\dot{M}_{\rm  euv} \propto d^{-1}$ \citep{1998apJ...499..758J}. However, if we set the $t_{\rm  dur}\to \infty $, $H$ increases slower and slower as time $t$ increase linearly. Furthermore, considering that the lifetime of the disc is finite, typically $\le$ 10 Myr, $H$ can not be too large. As the distance increases, the mass loss rate originated by external EUV photo-evaporation decreases. When $\dot{M}_{\rm  ex,euv} \le \dot{M}_{\rm  c,euv}$, i.e. the EUV photo-evaporation of external star is lower than the central star, we can neglect the external star and truncated the time as flyby times. Then if we combine the equation \ref{3} and \ref{7}, we can get the criterion where the external EUV photo-evaporation is not significant:
\begin{equation} \label{Equation B13}
d \ge 1.13 \times 10^{7} \rm \left| \cos\left(\theta\right) \right| \left(\frac{\Phi_{\rm i}}{10^{49}}\right)^{1/2} AU 
\end{equation}
For simplification we set $inc=0.5\pi$. For low mass stars with $\Phi_{\rm i} \sim 10^{43}$ photons s$^{-1}$, $d\ge 1.13 \times 10^{4}$ AU can satisfy the criterion. therefore flyby with low mass stars if $q\ge 1.13 \times 10^{4}$ AU only lead to few mass loss. In most cases, when we set $t_{\rm  dur}=2$ Myr, Equation \ref{Equation B13} is satisfied when $t>t_{dur}$. While for extreme massive stars with $\Phi_{\rm i} \sim 10^{49}$ photons s$^{-1}$, in orbit of low eccentricity $e\le 10$ and large peri-center distance $q\ge 10^{6}$ AU, the flyby duration should be set as large as 10 Myr.

In this paper, we fixed the flyby duration $t_{\rm  dur}=2 $ Myr in both single and multiple flyby models and after the duration, external photo-evaporation is not significant in most of simulated cases. From table \ref{Table B1}, we can see the mass loss caused by external EUV photo-evaporation in two cases with $t_{\rm dur}=2 $ Myr and $t_{\rm dur}=10 $ Myr for $q=10^4$ AU. We can see in the case of $t_{\rm dur}=2 $ Myr, larger than 88\% of the mass loss has been calculated. Similar to table \ref{Table B1}, mass loss during flyby in 2 Myr is about 80\% of the total mass loss in 10 Myr when $q=10^{5}$ AU. While in the case of $q=10^{6}$ AU, it's not reasonable to assume that the photoevaporation only last 2 Myr during flyby, because the mass loss caused in 2 Myr flyby(external EUV case) is only 25\% of that in 10 Myr. Further more, when it comes to the cases with smaller eccentricity $e=10$, mass loss in 2 Myr due to flyby also dominates the total mass loss in 10 Myr if $q=10^4$ and $10^5$ AU. While only about 20\% of total mass loss due to flyby is calculated, if $q=10^{6}$ AU. The results is reasonable, because When $q=10^{4}$ or $10^{5}$ AU, the distance between host star and external star is approximately change one or two magnitude in 2 Myr, the disc loses its mass near pericenter in 2 Myr. While in the case of $q=10^{6}$ AU, the distance is changed slowly and the mass loss changes a little, therefore, we should assume the mass loss due to flyby always works until it decays significantly. And when estimating the mass loss caused by external EUV photo-evaporation (both in external EUV case and EUV/FUV case with $10^{4} \le q \le 10^{5}$ AU), a 2 Myr flyby assumption is suitable. Therefore, our assumption of flyby duration is reasonable to estimate the most photo-evaporation effect of external star. Thus the maximum $H_{\rm m}$ during flyby events is $H_{\rm m}$ according to Equation \ref{Equation B10} by taking $t=t_{\rm dur}/2 $. We can rewrite the equation \ref{Equation B12}. For instance, the external star with $T_{\rm  eff}=19000$ K, $q=10^{4}$ AU and $e=100$, leads to $H_{\rm  m} \sim H \sim 10$,
\begin{equation} \label{B14}
\begin{aligned}
D\left(H_{\rm  m} \right) \approx &C_{\rm  0}\sin(inc)\sqrt{\frac{q/1\rm  AU}{e \mu}}
\left[- \frac{H(t_{\rm dur}/2)}{e} + 
\frac{\pi}{2} 
\right]
\end{aligned}
\end{equation}

\begin{table} 
	\begin{center}
		\begin{tabular}{cccc}
			\hline
			$T_{\rm   eff}$ [K] & $\Delta M_{r,2Myr}$  & $\Delta M_{r,10Myr}$ & $\frac{\Delta M_{r,2Myr}}{\Delta M_{r,10Myr}}$\\
			\hline
			\cline{1-4}
			15000&9.60$\times 10^{-5}$&1.09$\times 10^{-4}$&88.2\% \\
			\cline{1-4}
			19000&1.17$\times 10^{-3}$ & 1.33$\times 10^{-3}$&88.3\%\\
			\cline{1-4}
			25000&6.18$\times 10^{-3}$&6.98$\times 10^{-3}$&88.5\%\\
			\cline{1-4}
			30000&2.76$\times 10^{-2}$&3.12$\times 10^{-2}$&88.6\% \\
			\cline{1-4}
			34000&7.94$\times 10^{-2}$&8.95$\times 10^{-2}$&88.8\%\\
			\cline{1-4}
			38000&1.48$\times 10^{-1}$&1.66$\times 10^{-1}$&88.9\%\\   
			\hline
		\end{tabular}
		\caption{Relative mass loss due to the different external EUV photo-evaporation in different flyby duration: 2 Myr and 10 Myr. The first column is the effective temperature of the external star. The second and third column are represent the relative mass loss due to the different external EUV photo-evaporation in different flyby duration. And the last column is the ratio of $\Delta M_{r,2Myr}$ to $\Delta M_{r,10Myr}$. Other parameters are set as follows: orbit eccentricity $e=100$, pericenter distance $q=10^{4}$ AU \label{Table B1}}
	\end{center}
\end{table}

\end{document}